\documentstyle[aps,eqsecnum]{revtex}
\begin{document} 
\draft 
\title{Post-Newtonian expansion of gravitational waves 
from a particle in circular orbits around a rotating black hole: 
Up to $O(v^8)$ beyond the quadrupole formula} 
\author{Hideyuki Tagoshi} 
\address{California Institute of Technology, 
Theoretical Astrophysics, Pasadena, CA 91125, \\ 
and National Astronomical Observatory, 
Mitaka, Tokyo 181, Japan } 
\author{Masaru Shibata, Takahiro Tanaka, 
Misao Sasaki} 
\address{Department of Earth and Space Science, Osaka University, 
Toyonaka, Osaka 560, Japan}
\date{draft January 1996}
\maketitle
\begin{center}
\large{\bf Abstract}
\end{center}
\begin{abstract}
Extending a method developed by Sasaki in the Schwarzschild case and 
by Shibata, Sasaki, Tagoshi, and Tanaka in the Kerr case, 
we calculate the post-Newtonian expansion of the gravitational 
wave luminosities from a test particle in circular orbit 
around a rotating black hole up to $O(v^8)$ beyond 
the quadrupole formula. The orbit of a test particle is restricted 
on the equatorial plane. 
We find that spin dependent terms appear in each post-Newtonian order, and 
that at $O(v^6)$ they have a significant 
effect on the orbital phase evolution of coalescing compact binaries. 
By comparing the post-Newtonian formula of the luminosity with 
numerical results we find that, for $30M\lesssim r \lesssim 100M$, 
the spin dependent terms at $O(v^6)$ and $O(v^7)$ improve the accuracy of the 
post-Newtonian formula significantly, 
but those at $O(v^8)$ do not improve. 
\end{abstract} 

\section{Introduction}

Among the possible sources of gravitational waves, 
coalescing compact binaries are considered to be 
the most promising candidates for detection by near-future, 
ground based laser interferometric 
detectors such as LIGO \cite{Ref:ligo}, 
VIRGO\cite{Ref:virgo}, GEO600, TAMA and AIGO. 
There are two reasons for this: first, we 
can expect sufficiently large 
amplitude of gravitational waves from these systems. 
Second, the estimated event rate, for neutron star binaries, 
is several/yr within 200Mpc \cite{Ref:phinney}. 
Furthermore, the observations of coalescing compact binaries are potentially  
important because they 
bring us new physical and astronomical information. 
They can be used to test general relativity \cite{Ref:willBD}, 
to measure cosmological parameters \cite{Ref:markovic} and 
neutron star radii. It may even be possible to obtain information 
about the equation of state of neutron stars\cite{Ref:three}. 
If a neutron star or a small black hole spirals into a massive 
black hole with mass $<300M_\odot$, the inspiral wave form 
will be detected by the above detectors. 
Such wave forms carry detailed information about the  
spacetime geometry around the black hole, and therefore 
may be used to test the black hole no hair theorem\cite{Ref:RFT}. 

When a gravitational wave signal is detected, 
matched filtering will be used to 
extract the binary's 
parameters (i.e. masses, spins, ets.)\cite{Ref:three}. 
In this method, the parameters are determined 
by cross-correlating the noisy signal from the detectors 
with theoretical templates. 
If the signal and the templates lose phase with each other by 
one cycle over $\sim 10^3 -10^4$ cycles 
as the waves sweep through the LIGO/VIRGO band, their 
cross correlation will be significantly reduced. 
This means that we need  to construct theoretical templates 
which are accurate to better than one cycle during entire sweep
through the LIGO/VIRGO band \cite{Ref:three}. 
If we have accurate templates, we can, in principle, determine the 
mass of the systems within 1\% error \cite{Ref:flanagan}. 
Thus, much effort has been expended to 
construct accurate theoretical templates \cite{Ref:will}. 

The standard method to calculate inspiraling wave forms 
from coalescing binaries 
is the post-Newtonian expansion of the Einstein equations, 
in which the orbital velocity $v$ of the binaries 
are assumed to be small compared to the speed of light. 
Since, for coalescing binaries, the orbital velocity is not 
so small when the frequency of gravitational waves is in LIGO/VIRGO
band, it is necessary to carry the post-Newtonian expansion 
up to extremely high order in $v$. A post-Newtonian wave generation 
formalism which can handle the high order 
calculation has been developed by Blanchet, Damour 
and Iyer\cite{Ref:BandD,Ref:DI}. 
Based on this formalism, calculations have been carried out 
up to post$^{5/2}$-Newtonian order, or $O(v^5)$ beyond the leading 
order quadrupole formula[12-20]. 
An another formalism is also developed up to $O(v^4)$ 
by Will and Wiseman\cite{Ref:BDIWW,Ref:WW} 
which is based on Epstein-Wagoner's formalism\cite{Ref:EW,Ref:WagonerW}. 

Although the post-Newtonian calculation technique will be developed 
and applied to the higher order calculation, 
it will become more difficult and complicated. 
Thus, it would be very helpful if we could have another 
reliable method to calculate the higher order post-Newtonian corrections. 
Recently the post-Newtonian expansion based on black hole 
perturbation formalism is developed. 
In this analysis, 
one considers gravitational waves from a particle of mass $\mu$ 
orbiting a black hole of mass $M$ when $\mu \ll M$. 
Although this method is restricted to the case when $\mu\ll M$, 
one can calculate very high order post-Newtonian corrections to 
gravitational waves using a relatively simple algorithm 
in contrast with the standard post-Newtonian analysis. 
This direction of research was first done analytically by 
Poisson  \cite{Ref:poisson} who worked 
to $O(v^3)$ and numerically by Cutler et.al. \cite{Ref:cutler}
to $O(v^5)$. Subsequently, 
a highly accurate numerical calculation was carried out by 
Tagoshi and Nakamura \cite{Ref:TN} to $O(v^8)$
in which they found the appearance of $\log v$ terms 
in the energy flux at $O(v^6)$ and at $O(v^8)$. 
They also clarified that the accuracy of the energy flux to at least
$O(v^6)$ is needed to construct template wave forms for coalescing 
binaries. 
Tagoshi and Sasaki\cite{Ref:TS}, using the 
formulation built up by Sasaki \cite{Ref:sasaki}, performed analytic 
calculations which 
confirmed the result of Tagoshi and Nakamura. 
These calculations were extended to a rotating black hole 
case by Shibata, Sasaki, Tagoshi and Tanaka(SSTT)\cite{Ref:SSTT} 
to $O(v^5)$. They 
calculated gravitational waves from a particle in circular orbit 
with small inclination from the equatorial plane
to see the effect of spin at high post-Newtonian orders. 
They found that the effect of spin on the orbital phase 
is important at $O(v^5)$ order 
when one of the stars is a rapidly rotating 
neutron star with its pulse period less than 2 ms or a rapidly 
rotating black hole with $q=J_{BH}/M^2\geq 0.2$. 
This analysis was extended to the case of slightly eccentric orbits by 
Tagoshi \cite{Ref:tagoshi}. 
The absorption of gravitational waves into the black hole horizon, 
appearing at $O(v^8)$, was also calculated by Poisson and 
Sasaki in the case when a test particle is in a circular orbit around a 
Schwarzschild black hole\cite{Ref:PoSa}. 

In this paper, we extend these analyses in the rotating black hole case
to $O(v^8)$ order. Once again, the calculation is based on the formalism 
developed by Sasaki\cite{Ref:sasaki} to treat a Schwarzschild black hole. 
Based on the 
post-Newtonian expansion of the luminosity in the test particle limit 
when the central body is a Schwarzschild black hole 
(\cite{Ref:TN,Ref:TS}), Cutler and Flanagan \cite{Ref:CF} 
estimated that 
we will have to calculate post-Newtonian expansion 
of gravitational wave luminosity at least up to $O(v^6)$ 
in order to obtain the theoretical templates which 
cause less systematic errors than statistical errors for the 
LIGO detector. 
Further, in a previous paper \cite{Ref:SSTT}, we suggested that 
the effect of spin at $O(v^6)$
to the orbital phase of coalescing binaries wouldn't be negligible 
if spin of the black hole was large (i.e. $|q| \sim 1$). 
Also, the perturbation study can provide accurate templates 
for binaries with $M \gg \mu$(for example, binaries of $100M_{\odot}$ 
black hole-$1.4M_{\odot}$ neutron star). Since 
LIGO and VIRGO will be able to detect gravitational wave signals 
from binaries with masses 
less than $\sim 300 M_{\odot}$, it is important to construct templates for 
such binaries. The frequency of gravitational waves from such a massive 
binary, however,  comes into the frequency band for LIGO and VIRGO at 
$r/M \sim 16(100M_{\odot}/M)^{2/3}$, i.e., highly relativistic region. 
We do not know whether 
the convergence property of the post-Newtonian approximation is good or not 
in such a highly relativistic motion. Hence, it is an 
urgent problem to clarify 
at what point the convergence property of the post-Newtonian 
expansion is good. For these purposes, 
we study the effect of spin beyond $O(v^6)$ order in this paper. 

The paper is organized as follows. 
In section 2, we present the basic formalism to perform the post-Newtonian
expansion in our perturbative approach. 
First we perform the post-Newtonian expansion of the 
Teukolsky radial function using the Sasaki-Nakamura equation. 
We also show the post-Newtonian expansion of 
the angular equation, which is given in Appendix F.
In section 3, we first describe the post-Newtonian expansion 
of the source terms. 
We consider circular orbits in the equatorial plane 
around a Kerr black hole. 
Then the gravitational wave luminosities to 
$O(v^8)$ beyond the quadrupole formula are derived. 
In section 4, we compare post-Newtonian formulas
with numerical data which gives the exact value of gravitational wave 
luminosity and investigate the convergence property 
of the post-Newtonian expansion. 
Section 5 is devoted to summary and discussion. 

Throughout this paper we use the units of $c=G=1$. 

\section{General formulation}
\subsection{The Teukolsky equation}
We consider the case when a test particle of mass $\mu$ travels 
in a circular orbit around a Kerr black hole of mass 
$M\gg\mu$. 
We follow the notation used by SSTT\cite{Ref:SSTT}, but for 
definiteness, 
we recapitulate necessary formulas and definitions.

To calculate gravitational radiation from a particle orbiting 
a Kerr black hole, we start with 
the Teukolsky equation\cite{Ref:Teu,Ref:breuer}.  We focus on the
radiation going out to infinity described by 
the fourth Newman-Penrose quantity, $\psi_4$\ \cite{Ref:New}, 
which may be expressed as
\begin{equation}
\psi_4=(r-ia\cos\theta)^{-4}\int d\omega
 e^{-i\omega t}\sum_{\ell,m}
{e^{im\varphi}\over\sqrt{2\pi}}\,
{}_{-2}S^{a\omega}_{\ell m}(\theta)R_{\ell m\omega}(r),
\label{eq:psifour}
\end{equation}
where $_{-2}S_{\ell m}^{a\omega}$ is the spheroidal harmonic function of 
spin weight $s=-2$, which is normalized as
\begin{equation}
\int_0^{\pi} |_{-2}S_{\ell m}^{a\omega}|^2 \sin\theta d\theta=1.
\label{eq:Snorm}
\end{equation}
The radial function $R_{\ell m\omega}(r)$ obeys the Teukolsky equation
with spin weight $s=-2$, 
\begin{equation}
\Delta^2 {d \over dr}\Bigl({1 \over \Delta}{dR_{\ell m\omega} \over dr}
\Bigr) 
-V(r)R_{\ell m\omega}=T_{\ell m\omega}(r),
\end{equation}
where $T_{\ell m\omega}(r)$ 
is the source term whose explicit form will be shown later, and 
$\Delta=r^2-2Mr+a^2$. The potential $V(r)$ is given by
\begin{equation}
V(r)=-{K^2+4i(r-M)K \over \Delta}+8i\omega r+\lambda,
\end{equation}
where $K=(r^2+a^2)\omega-ma$ and $\lambda$ is the eigenvalue 
of $_{-2}S_{\ell m}^{a\omega}$. 

The solution of the Teukolsky equation at infinity 
($r\rightarrow\infty$) is expressed as
\begin{equation}
R_{\ell m\omega}(r)
 \rightarrow{r^3e^{i\omega r^*} \over 2i\omega B^{{\rm in}}_{\ell m\omega}}
\int^{\infty}_{r_+}dr'{T_{\ell m\omega}(r') R_{\ell m\omega}^{{\rm in}}(r') 
\over\Delta^{2}(r')}
\equiv \tilde Z_{\ell m\omega}r^3e^{i\omega r^*},
\label{eq:Rinfty}
\end{equation}
where $r_+=M+\sqrt{M^2-a^2}$ denotes the radius of the event horizon
and $R^{{\rm in}}_{\ell m\omega}$ is the homogeneous solution
which satisfies the ingoing-wave boundary condition at horizon,
\begin{equation}
R^{{\rm in}}_{\ell m\omega} \rightarrow\cases{
D_{\ell m\omega}\Delta^2  e^{-ikr^*}\,&
for $r^*\rightarrow -\infty$\cr
r^3 B_{\ell m\omega}^{{\rm  out}}e^{i\omega r^*}+
r^{-1}B_{\ell m\omega}^{{\rm in}} e^{-i\omega r^*}\,&
for $r^*\rightarrow +\infty,$ \cr}
\label{eq:Rin}
\end{equation}
where $k=\omega-ma/2Mr_+$ and
$r^*$ is the tortoise coordinate defined by 
\begin{equation}
{dr^* \over dr}={r^2+a^2 \over \Delta}.
\end{equation}
For definiteness, we fix the integration constant such that
$r^*$ is given explicitly by
\begin{eqnarray}
r^{*}&=&\int {dr^*\over dr}dr \nonumber\\
&=&r+{2Mr_{+}\over {r_{+}-r_{-}}}\ln{{r-r_{+}}\over 2M}
-{2Mr_{-}\over {r_{+}-r_{-}}}\ln{{r-r_{-}}\over 2M},
\label{eq:rst}\\
\nonumber
\end{eqnarray}
where $r_{\pm}=M\pm \sqrt{M^2-a^2}$.

\subsection{Post-Newtonian expansion of the homogeneous solution}
In the previous papers \cite{Ref:sasaki,Ref:SSTT}, 
the post-Newtonian expansion of the homogeneous
solution was performed to $O(\epsilon^2)$ 
in the Schwarzschild case and $O(\epsilon)$ in the Kerr case, 
where $\epsilon\equiv 2M\omega$. 
In this section, we extend those methods, performing the expansion of
homogeneous solutions up to $O(\epsilon^2)$. 

In order to calculate gravitational waves emitted to infinity from
a particle in a circular orbit,
we need to know the explicit form of the source term $T_{\ell m\omega}(r)$,
which has support only at $r=r_0$ where $r_0$ is orbital radius 
in the Boyer-Lindquist coordinate, 
the ingoing-wave Teukolsky function $R^{{\rm in}}_{\ell m\omega}(r)$ 
at $r=r_0$, 
and its incident amplitude $B^{{\rm in}}_{\ell m\omega}$ at infinity.
We consider the expansion of these quantities in terms of a small parameter, 
$v^2 \equiv M/r_0$. 
In addition, we need to expand those quantity
in terms of $\epsilon\equiv 2M\omega$ since $\omega\sim O(\Omega)$ 
where $\Omega$ is the orbital angular 
velocity of the particle and $M\omega\sim O(v^3)$. 
In the case of a Kerr black hole, other combination of 
parameters $a\omega$ appears in the Teukolsky equation. 
We define $q\equiv a/M$ and we have $a\omega$ $=q \epsilon/2\sim 
O(\epsilon^3)$. 

First we perform the expansion of the spheroidal harmonics 
$_{-2}S^{a\omega}_{\ell m}$ and their 
eigenvalues $\lambda$ in terms of $a\omega$. 
Since $a\omega=O(v^3)$, we have to calculate 
$_{-2}S^{a\omega}_{\ell m}$ and $\lambda$ up to $O((a\omega)^2)$. 
The eigenvalue $\lambda$ has already been evaluated 
up to $O((a\omega)^2)$ in a 
previous paper\cite{Ref:SSTT}. We calculate the expansion of 
$_{-2}S^{a\omega}_{\ell m}$ at $O((a\omega)^2)$ in the Appendix F. 
As a result, the spheroidal harmonics 
$_{-2}S^{a\omega}_{\ell m}$ are given by
\begin{equation} 
{}_{-2}S_{\ell m}^{a\omega}
={}_{-2}P_{\ell m}+a\omega S_{\ell m}^{(1)}
+(a\omega)^2 S_{\ell m}^{(2)}
+O((a\omega)^3),
\end{equation}
where ${}_{-2}P_{\ell m}$ are the spherical harmonics of spin weight
$s=-2$ \cite{Ref:spin} and 
\begin{equation}
S_{\ell m}^{(1)}=\sum_{\ell'} c_{\ell m}^{\ell'}~{}_{-2}P_{\ell'm}\,.
\end{equation}
Here $c_{\ell m}^{\ell'}$ are non-zero only for $\ell'=\ell \pm 1$, 
explicitly
\begin{eqnarray*}
& &c_{\ell m}^{\ell+1}={2 \over (\ell+1)^2}\Bigl\lbrack 
{(\ell+3)(\ell-1)(\ell+m+1)(\ell-m+1) \over (2\ell+1)(2\ell+3)}
 \Bigr\rbrack^{1/2},\\
& &c_{\ell m}^{\ell-1}=-{2 \over \ell^2}\Bigl\lbrack 
{(\ell+2)(\ell-2)(\ell+m)(\ell-m) \over (2\ell+1)(2\ell-1)} 
\Bigr\rbrack^{1/2}.\\
\end{eqnarray*}
$S_{\ell m}^{(2)}$ is given by
\begin{equation}
S_{\ell m}^{(2)}=\sum_{\ell'} d_{\ell m}^{\ell'}~{}_{-2}P_{\ell'm}\,,
\end{equation}
where the non-zero components of $d_{\ell m}^{\ell'}$ are given by
\begin{equation}
d_{\ell m}^{\ell}=-{1\over 2}\left( \left(c_{\ell m}^{\ell+1}\right)^2+
\left(c_{\ell m}^{\ell-1}\right)^2\right)\,,
\end{equation}
for any $\ell$, by 
\begin{eqnarray*}
d_{\ell m}^{\ell+1}&=&{m\over 324\sqrt{7}}(3-m)^{1/2}(3+m)^{1/2},\\
d_{\ell m}^{\ell+2}&=&{11\over 1764\sqrt{3}}(3-m)^{1/2}(3+m)^{1/2}
(4-m)^{1/2}(4+m)^{1/2},\\
\end{eqnarray*}
for $\ell=2$, and by 
\begin{eqnarray*}
d_{\ell m}^{\ell+1}&=&{m\over 120\sqrt{21}}(4-m)^{1/2}(4+m)^{1/2},\\
d_{\ell m}^{\ell+2}&=&{1\over 180\sqrt{11}}(4-m)^{1/2}(4+m)^{1/2}
(5-m)^{1/2}(5+m)^{1/2},\\
d_{\ell m}^{\ell-1}&=&-{m\over 324\sqrt{7}}(3-m)^{1/2}(3+m)^{1/2},\\
\end{eqnarray*}
for $\ell=3$. We don't need $S_{\ell m}^{(2)}$ for $\ell=4$ in this 
paper.

The eigenvalue $\lambda$ is given by 
\begin{equation}
\lambda=\lambda_0+a\omega\lambda_1+a^2\omega^2 \lambda_2
+O((a\omega)^3)
\end{equation}
where $\lambda_0=(\ell-1)(\ell+2)$, 
$\lambda_1=-2m(\ell^2+\ell+4)/(\ell^2+\ell)$ and
\begin{equation}
\lambda_2=-2(\ell+1)(c_{\ell m}^{\ell+1})^2+2\ell(c_{\ell m}^{\ell-1})^2
+{2\over 3}-{2 \over 3}{(\ell+4)(\ell-3)(\ell^2+\ell-3m^2) \over 
\ell(\ell+1)(2\ell+3)(2\ell-1)}\,.
\end{equation}

Next we calculate the homogeneous solution $R^{{\rm in}}_{\ell m\omega}$. 
Here we only consider the case when $\omega>0$. We must treat the case 
$\omega\leq 0$ separately. 
The Teukolsky equation is transformed into the 
Sasaki-Nakamura equation \cite{Ref:SN},
which is given by 
\begin{equation}
\Bigl[{d^2 \over dr^{*2}}-F(r){d \over dr^{*}}-U(r)\Bigr]
X_{\ell m\omega}=0. \label{eq:sneq}
\end{equation}
The explicit forms of $F(r)$ and $U(r)$ are given in the 
Appendix A. The relation between $R_{\ell m\omega}$
and $X_{\ell m\omega}$ is
\begin{equation}
R_{\ell m\omega}={1 \over \eta}
\Bigl\{\Bigl(\alpha+{\beta_{,r} \over \Delta}\Bigr)\chi_{\ell m\omega}
-{\beta \over \Delta}\chi_{\ell m\omega,r} \Bigr\},\label{eq:xtor}
\end{equation}
where $\chi_{\ell m\omega}=X_{\ell m\omega} \Delta/(r^2+a^2)^{1/2}$, and 
the functions $\alpha$, $\beta$ and $\eta$ are shown in Appendix A.
Conversely, we can express $X_{\ell m\omega}$ in terms of 
$R_{\ell m\omega}$ as
\begin{equation}
X_{\ell m\omega}=(r^2+a^2)^{1/2}\,r^2\,J_{-}J_{-}
\left[{1\over r^2}R_{\ell m\omega}\right], \label{eq:rtox}
\end{equation}
where $J_{-}=(d/dr)-i(K/\Delta)$. 
Then the asymptotic behavior of 
the ingoing-wave solution $X_{\ell m\omega}^{{\rm in}}$ which corresponds
to Eq.(\ref{eq:Rin}) is 
\begin{equation}
X_{\ell m\omega}^{{\rm in}}\rightarrow\left\{
\begin{array}{ll}
A_{\ell m\omega}^{{\rm out}}e^{i\omega r^*}+
A_{\ell m\omega}^{{\rm in}} e^{-i\omega r^*}\,& 
{\rm for}~r^* \rightarrow \infty\\
C_{\ell m\omega} e^{-ik r^*}\, & {\rm for}~
r^* \rightarrow -\infty. 
\end{array}
\right.\label{eq:xinasy}
\end{equation}
The coefficient 
$A_{\ell m\omega}^{{\rm in}}$, $A_{\ell m\omega}^{{\rm out}}$ and 
$C_{\ell m\omega}$ are respectively related to 
$B_{\ell m\omega}^{{\rm in}}$, $B_{\ell m\omega}^{{\rm out}}$ and 
$D_{\ell m\omega}$, defined in Eq.(\ref{eq:Rin}), by 
\begin{eqnarray}
B_{\ell m\omega}^{{\rm in}}&=&-{1\over 4\omega^2}A_{\ell m\omega}^{{\rm in}}, 
\nonumber\\
B_{\ell m\omega}^{{\rm out}}&=&-{4\omega^2\over c_0}
A_{\ell m\omega}^{{\rm out}}, 
\label{eq:ainbin}\\
D_{\ell m\omega}&=&{1\over d_{\ell m\omega}}C_{\ell m\omega}, 
\nonumber\\
\nonumber
\end{eqnarray}
where $c_0$ is given in Eq.(A.3) of Appendix A and
\begin{eqnarray*}
d_{\ell m\omega}&=&\sqrt{2Mr_+}[(8-24iM\omega-16M^2\omega^2)r_{+}^2 \\
& &+(12iam-16M+16amM\omega+24iM^2\omega)r_{+}
-4a^2m^2-12iamM+8M^2]. \\
\end{eqnarray*}

Now we introduce the variable $z=\omega r$ and
\begin{eqnarray}
z^*&=&z+\epsilon\left[{z_{+}\over {z_{+}-z_{-}}}\ln(z-z_+)
-{z_{-}\over {z_{+}-z_{-}}}\ln(z-z_-)\right]\nonumber\\
&=&\omega r^*+\epsilon\ln\epsilon\,, \label{eq:zst}\\
\nonumber
\end{eqnarray}
where $z_\pm=\omega r_\pm$. 
To solve $X^{{\rm in}}_{\ell m\omega}$ by expanding it in terms of $\epsilon$,
we set
\begin{equation}
X^{{\rm in}}_{\ell m\omega}=\sqrt{z^2+a^2\omega^2}\xi_{\ell m}(z)
\exp\left(-i\phi(z)\right), 
\label{eq:xixi}
\end{equation}
where 
\begin{eqnarray}
\phi(z) &=&\int dr\left({K\over \Delta}-\omega\right)\nonumber\\
&=&z^*-z-{\epsilon\over 2}mq{1\over {z_{+}-z_{-}}}
\ln{{z-z_{+}}\over {z-z_{-}}}\,,\label{eq:phase}\\
\nonumber
\end{eqnarray}
which generalizes the phase function $\omega(r^*-r)$ of
the Schwarzschild case.
This prescription makes it easy to implement the ingoing-wave
boundary condition on $X^{{\rm in}}_{\ell m\omega}$.

Inserting Eq.($\ref{eq:xixi}$) into Eq.($\ref{eq:sneq}$) 
and expanding it in powers of $\epsilon=2M\omega$, we obtain
\begin{equation}
L^{(0)}[\xi_{\ell m}]=\epsilon L^{(1)}[\xi_{\ell m}]
+\epsilon Q^{(1)}[\xi_{\ell m}]+\epsilon^2 Q^{(2)}[\xi_{\ell m}]
+\epsilon^3 Q^{(3)}[\xi_{\ell m}]+\epsilon^4 Q^{(4)}[\xi_{\ell m}]
+O(\epsilon^5), \label{eq:eps}
\end{equation}
where $L^{(0)}$, $L^{(1)}$, $Q^{(1)}$ and $Q^{(2)}$ are differential 
operators given by
\begin{eqnarray}
L^{(0)}&=&{d^2\over dz^2}+{2\over z}{d\over dz}+\left(1-{\ell(\ell+1)
\over z^2}\right), \\
L^{(1)}&=&{1\over z}{d^2\over dz^2}
+\left( {{1}\over z^2}
+{2i\over z}\right){d\over dz}
-\left( {{4}\over {z^3}}-{i\over z^2}
+{1\over z}\right), \\
Q^{(1)}&=&{{iq\lambda_1}\over 2z^2}{d\over dz}
-{{4imq}\over {l(l+1)z^3}}, \\
\nonumber
\end{eqnarray} 
and $Q^{(2)}$, $Q^{(3)}$ and $Q^{(4)}$ are 
given in Appendix C. Note that the real part of
$Q^{(1)}$ vanishes when we insert  the expression for $\lambda_1$. 
There are $\lambda_3$ or $\lambda_4$ in the formulas for 
$Q^{(3)}$ and $Q^{(4)}$. However, it is straghtforward to show that 
both $\lambda_3$ and $\lambda_4$ do not influence the results 
in this paper. 

By expanding $\xi_{\ell m}$ in terms of $\epsilon$ as
\begin{equation}
\xi_{\ell m}=\sum_{n=0}^{\infty}\epsilon^n \xi^{(n)}_{\ell m}(z), 
\label{eq:expa}
\end{equation}
we obtain from Eq.(\ref{eq:eps}) the iterative equations, 
\begin{eqnarray}
& &L^{(0)}[\xi^{(0)}_{\ell m}]=0, \label{eq:zeroji}\\
& &L^{(0)}[\xi^{(1)}_{\ell m}]=L^{(1)}[\xi^{(0)}_{\ell m}]
+Q^{(1)}[\xi^{(0)}_{\ell m}]\equiv W^{(1)}_{\ell m}, 
\label{eq:first}\\
& &L^{(0)}[\xi^{(2)}_{\ell m}]=L^{(1)}[\xi^{(1)}_{\ell m}]
+Q^{(1)}[\xi^{(1)}_{\ell m}]+Q^{(2)}[\xi^{(0)}_{\ell m}]
\equiv W^{(2)}_{\ell m}. \label{eq:second}\\
& &L^{(0)}[\xi^{(3)}_{\ell m}]=L^{(1)}[\xi^{(2)}_{\ell m}]
+Q^{(1)}[\xi^{(2)}_{\ell m}]+Q^{(2)}[\xi^{(1)}_{\ell m}]
+Q^{(3)}[\xi^{(0)}_{\ell m}]
\equiv W^{(3)}_{\ell m}. \label{eq:third}\\
& &L^{(0)}[\xi^{(4)}_{\ell m}]=L^{(1)}[\xi^{(3)}_{\ell m}]
+Q^{(1)}[\xi^{(3)}_{\ell m}]+Q^{(2)}[\xi^{(2)}_{\ell m}]
+Q^{(3)}[\xi^{(1)}_{\ell m}]+Q^{(4)}[\xi^{(0)}_{\ell m}]
\equiv W^{(4)}_{\ell m}. \label{eq:forth}\\
\nonumber
\end{eqnarray}
The general solution to Eq.(\ref{eq:zeroji}) 
is immediately obtained as 
\begin{equation}
\xi^{(0)}_{\ell m}=\alpha_\ell^{(0)} j_\ell+\beta_\ell^{(0)} n_\ell, 
\label{eq:xizero}
\end{equation} 
where $j_{\ell}$ and $n_{\ell}$ are the usual spherical Bessel functions. 
As we discuss later, the boundary condition for $n\leq 2$ is that 
$\xi^{(n)}_{\ell m}$ is regular at $z=0$. 
Hence $\beta_\ell^{(0)}=0$ and we set $\alpha_\ell^{(0)}=1$ for
convenience.

To calculate $\xi^{(n)}_{\ell m}$ for $n\geq 1$, we rewrite
Eqs.(\ref{eq:first}) $-$ (\ref{eq:forth}) in the indefinite
integral form by using the spherical Bessel functions as
\begin{equation}
\xi^{(n)}_{\ell m}=n_\ell\int^z dz z^2 j_\ell W^{(n)}_{\ell m}-
j_\ell\int^z dz z^2 n_\ell W^{(n)}_{\ell m}\quad(n=1,2). 
\label{eq:green}
\end{equation}
The calculation for $n=1$ was done in a previous paper\cite{Ref:SSTT}
and we have
\begin{eqnarray}
\xi^{(1)}_{\ell m} &=&\alpha^{(1)}_{\ell }j_\ell 
+{(\ell -1)(\ell +3)\over 2(\ell +1)(2\ell +1)}
j_{\ell +1}-{{\ell ^2-4}\over 2\ell (2\ell+1)}j_{\ell-1} \nonumber\\
&+&z^2(n_\ell j_0-j_\ell n_0)j_0
+\sum_{k=1}^{\ell-1}\left({1\over k}+{1\over k+1}\right)
z^2(n_\ell j_k-j_\ell n_k)j_k \cr
&+&n_\ell({\rm Ci} (2z)-\gamma-\ln 2z )-j_\ell{\rm Si} (2z) 
+ij_\ell\ln z \nonumber\\
&+&{imq\over 2} \left({\ell^2+4}\over \ell^2(2\ell+1)\right)j_{\ell-1}
+{imq\over 2} \left({(\ell+1)^2+4}\over {(\ell+1)^2(2\ell+1)}\right)
j_{\ell+1},\label{eq:xsi}\\
\nonumber
\end{eqnarray}
where ${\rm Ci}(x)=-\int^{\infty}_x dt\cos t/t$ and 
${{\rm Si}(x)}=\int^x_0 dt\sin t/t$ are cosine and sine integral functions, 
$\gamma$ is the Euler constant, and
$\alpha^{(1)}_\ell$ is an integration constant which represents the 
arbitrariness of the normalization of 
$X_{\ell m\omega}^{{\rm in}}$. We set $\alpha^{(1)}_\ell=0$ for simplicity. 

Next we consider $\xi^{(2)}_{\ell m}$. From Eqs.(\ref{eq:green}) and 
(\ref{eq:xsi}), and by using formulas in the paper\cite{Ref:sasaki},
we obtain $\xi^{(2)}_{\ell m}$ as 
\begin{equation}
\xi^{(2)}_{\ell m}=f^{(2)}_{\ell}+ig^{(2)}_{\ell}+k^{(2)}_{\ell m}(q)
+\alpha^{(2)}_{\ell m} j_\ell+\beta^{(2)}_{\ell m} n_\ell\,,
\end{equation}
where $f^{(2)}_{\ell}$ and $g^{(2)}_{\ell}$ are the real and imaginary part
of $\xi^{(2)}_{\ell m}$ in the Schwarzschild case respectively,
$k^{(2)}_{\ell m}(q)$ exists only in Kerr case and 
$\alpha^{(2)}_{\ell m}$ and $\beta^{(2)}_{\ell m}$ are arbitrary constants.
The explicit forms of $f^{(2)}_{\ell}$ and $g^{(2)}_{\ell}$ are given 
in a previous paper\cite{Ref:sasaki}. 
The term $k^{(2)}_{\ell m}(q)$ is given for $\ell=2$ by 
\begin{eqnarray}
k^{(2)}_{2 m}&=&
{{191\,i}\over {180}}\,m\,q\,j_0 - {{{m^2}\,{q^2}\,j_0}\over {30}} - 
  {{m\,q\,j_1}\over {10}} \nonumber\\
& &- {{68\,i}\over {63}}\,m\,q j_2 - 
     {{{q^2}}\over {392}} j_2 - {{73\,{m^2}\,{q^2}}\over {1764}}\,j_2 
+ {{7\,m\,q\,j_3}\over {180}} - {i\over {72}}\,{q^2}\,j_3 + 
  {i\over {324}}\,{m^2}\,{q^2}\,j_3 \nonumber\\
& &+ {{11\,i}\over {420}}\,m\,q\,j_4 - {{{q^2}\,j_4}\over {392}} - 
  {{71\,{m^2}\,{q^2}\,j_4}\over {8820}} 
+  {{13\,i}\over 6}\,m\,q\,n_1 \nonumber\\
& &+ \left( -{{ m\,q\,j_1 }\over 5} - 
     {{13\,m\,q\,j_3}\over {90}} \right) \,\ln z 
+   \left( {{-i}\over 5}\,m\,q\,j_1 - 
     {{13\,i}\over {90}}\,m\,q\,j_3 \right) {\rm S}(z) \nonumber\\
& &+ \left( {i\over 5}\,m\,q\,n_1 + {{13\,i}\over {90}}\,m\,q\,n_3 
\right){\rm C}(z), 
\end{eqnarray}
and for $\ell=3$
\begin{eqnarray}
k^{(2)}_{3 m}&=&
{{3527\,i}\over {840}}\,m\,q\,j_1 - {{2\,{m^2}\,{q^2}\,j_1}\over {315}} - 
  {{m\,q\,j_2}\over {36}} - {{5\,i}\over {504}}\,{q^2}\,j_2 + 
  {{5\,i}\over {2268}}\,{m^2}\,{q^2}\,j_2 \nonumber\\
& &- {{379\,i}\over {360}}\,m\,q\,j_3 - 
  {{{q^2}\,j_3}\over {360}} - {{7\,{m^2}\,{q^2}\,j_3}\over {720}} + 
  {{3\,m\,q\,j_4}\over {160}} - {i\over {140}}\,{q^2}\,j_4 + 
  {i\over {1120}}\,{m^2}\,{q^2}\,j_4 \nonumber\\ 
& &+ {{97\,i}\over {5040}}\,m\,q\,j_5 - {{{q^2}\,j_5}\over {360}} - 
  {{17\,{m^2}\,{q^2}\,j_5}\over {5040}} 
- {{103\,i}\over {48}}\,m\,q\,n_0 + 
  {{25\,i}\over 8}\,m\,q\,n_2 \nonumber\\ 
& &- \left( {{13\,m\,q\,j_2}\over {126}} + 
  {{5\,m\,q\,j_4}\over {56}} \right) \,\ln z
+ \left( {{-13\,i}\over {126}}\,m\,q\,j_2 - 
     {{5\,i}\over {56}}\,m\,q\,j_4 \right) {\rm S}(z) \nonumber\\ 
& &+ \left( {{13\,i}\over {126}}\,m\,q\,n_2 + 
     {{5\,i}\over {56}}\,m\,q\,n_4 \right) {\rm C}(z), 
\end{eqnarray} 
where ${\rm C}(z)={\rm Ci} 2z - \gamma - \ln 2z$ and
${\rm S}(z)={\rm Si} 2z$. 
Note that to obtain above two formulas, we have added 
terms proportional to $j_{\ell}$ to simplify the 
formulas of $A^{{\rm in}}_{\ell m\omega}$ below. 
As noted previously, 
the source term $T_{\ell m\omega}$ has support only at $r=r_0$ and
$\omega r_0=O(v)$. 
Hence we only need $X^{{\rm in}}_{\ell m\omega}$ at
$z=O(v)\ll 1$ to evaluate the source integral, apart from the value
of the incident amplitude $A^{{\rm in}}_{\ell m\omega}$.
Hence the post-Newtonian expansion 
of $X^{{\rm in}}_{\ell m\omega}$ corresponds to the expansion 
not only in terms of $\epsilon=O(v^3)$, but also 
$z$ by assuming $\epsilon\ll z\ll 1$.
In order to evaluate the gravitational 
wave luminosity to $O(v^8)$ beyond the leading order, 
we must calculate the series expansion of $\xi^{(n)}_{\ell m}$ in powers
of $z$ for $n=0$ to $\ell=6$, for $n=1$ to $\ell=5$, 
for $n=2$ to $\ell=4$, for $n=3$ to $\ell=3$ and 
for $n=4$ to $\ell=2$ (See Appendix C of SSTT). 

When we evaluate $A^{{\rm in}}_{\ell m\omega}$, we examine the asymptotic 
behavior of $\xi^{(n)}_{\ell m}$ at infinity. Since the accuracy 
of $A^{{\rm in}}_{\ell m\omega}$ we need is $O(\epsilon^2)$, we don't have to
calculate $\xi^{(3)}_{\ell m}$ and $\xi^{(4)}_{\ell m}$ 
in closed analytic form. We need only
the series expansion formulas for $\xi^{(3)}_{\ell m}$ and
$\xi^{(4)}_{\ell m}$ around $z=0$, 
which is easily obtained by Eq.(\ref{eq:green}). 
Inserting $\xi^{(n)}_{\ell m}$ into Eq.(\ref{eq:xixi}) and expanding it
by $z$ and $\epsilon$ assuming $\epsilon\ll z\ll 1$, we obtain
\begin{eqnarray}
\xi^{(3)}_{2 m}&=&
{{{-{q^2}}\over {30z}} - {i\over {30z}}\,m\,{q^3}}
+ {{-i}\over {30}} + {{7\,m\,q}\over {180}} - 
  {i\over {60}}\,{m^2}\,{q^2} + 
  {{m\,{q^3}}\over {36}} - 
  {{{m^3}\,{q^3}}\over {90}} \nonumber\\
& &
- {{m\,q\,\ln z}\over {30}} - 
  {i\over {30}}\,{m^2}\,{q^2}\,\ln z \nonumber\\
& &
+  z\,\left( {{319}\over {6300}} + 
     {{100637\,i}\over {441000}}\,m\,q - 
     {{{q^2}}\over {180}} 
+    {{17\,{m^2}\,{q^2}}\over {1134}} + 
     {{83\,i}\over {5880}}\,m\,{q^3} 
\right.\nonumber\\
& &\left.
-    {{61\,i}\over {13230}}\,{m^3}\,{q^3} + 
     {{\ln z}\over {15}} - 
     {{106\,i}\over {1575}}\,m\,q\,\ln z - 
     {i\over {30}}\,m\,q\,{{\ln z}^2} \right) 
\nonumber\\
& &
+O(z^2)
+\alpha_{2 m}^{(3)} j_2+\beta_{2 m}^{(3)} n_2, \\
\xi^{(4)}_{2 m}&=&{{{q^4}}\over {80\,{z^2}}}
+O(z^{-1})
+\alpha_{2 m}^{(4)} j_2+\beta_{2 m}^{(4)} n_2,\\
\xi^{(3)}_{3 m}&=&
{{-i}\over {1260}}\,m\,q + 
  {{{m^2}\,{q^2}}\over {1890}} - 
  {i\over {1260}}\,m\,{q^3} - 
  {i\over {3780}}\,{m^3}\,{q^3}
 +O(z) \nonumber\\
& &
+\alpha_{3 m}^{(3)} j_3+\beta_{3 m}^{(3)} n_3. 
\end{eqnarray} 
The boundary condition of $\xi^{(n)}_{\ell m}$ that correctly
represent the boundary condition of $X^{{\rm in}}_{\ell m\omega}$ 
(Eq.(\ref{eq:xinasy})) is that 
$z \xi^{(n)}_{\ell m}$ must be no more singular than $z^{(\ell+1-n)}$ at
$z\rightarrow 0$. Since we need $\xi^{(n)}_{\ell m}$ only up to 
$n=4$, we set $\beta^{(n)}_{\ell m}=0$ for all of $\ell$ and $n$ in this 
paper. 
As for $\alpha^{(n)}_{\ell m}$, they still remain arbitrary and we set 
$\alpha^{(n)}_{\ell m}=0$ for all of $\ell$, $m$ and $n=3,4$. 

Inserting $\xi^{(n)}_{\ell m}$ into Eq.(\ref{eq:phase}) and expanding it
in terms of $\epsilon=2M\omega$, we obtain $X^{{\rm in}}_{\ell m\omega}$ 
which are shown in Appendix C. 
Using the transformation of Eq.(\ref{eq:xtor}), we obtain 
$R^{{\rm in}}_{\ell m\omega}$ which are also shown in Appendix D. 

Next, we consider $A^{{\rm in}}_{\ell m\omega}$ at $O(\epsilon^2)$. 
Using the relation $j_{\ell+1}\sim$ $-j_{\ell-1}\sim$ 
$(-1)^{\ell+n}n_{2n-\ell}$ at $z\sim\infty$, etc., 
we obtain the asymptotic behavior of 
$\xi^{(1)}_{\ell m}$ and $\xi^{(2)}_{\ell m}$
at $z\sim\infty$ as 
\begin{eqnarray}
\xi^{(1)}_{\ell m}&\sim& p^{(1)}_{\ell m} j_\ell+(q^{(1)}_{\ell m}
-\ln z)n_\ell+i j_\ell \ln z, \\
\xi^{(2)}_{\ell m}&\sim& 
\left(p^{(2)}_{\ell m}+q^{(1)}_{\ell m} \ln z-(\ln z)^2\right) j_\ell
+(q^{(2)}_{\ell m}-p^{(1)}_{\ell m} \ln z) n_\ell
\nonumber\\
& &
+i p^{(1)}_{\ell m}j_\ell \ln z+i(q^{(1)}_{\ell m}-\ln z) 
n_\ell \ln z
\end{eqnarray}
where
\begin{eqnarray}
p^{(1)}_{\ell m}&=&-{\pi\over 2}, \\
q^{(1)}_{\ell m}&=&{1\over 2}\left[\psi(\ell)+\psi(\ell+1)
+{{(\ell-1)(\ell+3)}\over \ell(\ell+1)}\right]-\ln 2
-{{2imq}\over {\ell^2(\ell+1)^2}}\,, \label{eq:qichi}\\
\psi(\ell)&=&\sum^{\ell-1}_{k=1}{1\over k}-\gamma, 
\end{eqnarray}
for any $\ell$ and 
\begin{eqnarray}
p^{(2)}_{2 m}&=&
{{457\,\gamma}\over {210}} - 
  {{{{\gamma}^2}}\over 2} + 
  {{5\,{{\pi }^2}}\over {24}} - 
  {i\over {18}}\,\gamma\,m\,q + 
  {{457\,\ln 2}\over {210}} \nonumber\\
& &
- \gamma\,\ln 2 - 
  {i\over {18}}\,m\,q\,\ln 2 - 
  {{{({\ln 2})^2}}\over 2}, \\
q^{(2)}_{2 m}&=&
     {{-457\,\pi }\over {420}} + 
     {{\gamma\,\pi }\over 2} + 
     {{5\,m\,q}\over {36}} + 
     {i\over {36}}\,m\,\pi \,q - 
     {i\over {72}}\,{q^2} + 
     {i\over {324}}\,{m^2}\,{q^2} + 
     {{\pi \,\ln 2}\over 2}, \\
p^{(2)}_{3 m}&=&
{{52\,\gamma}\over {21}} - {{{{\gamma}^2}}\over 2} + 
  {{5\,{{\pi }^2}}\over {24}} - {i\over {72}}\,\gamma\,m\,q + 
  {{52\,\ln 2}\over {21}} \nonumber\\
& &- \gamma\,\ln 2 - 
  {i\over {72}}\,m\,q\,\ln 2 - {{{({\ln 2})^2}}\over 2}, \\
q^{(2)}_{3 m}&=&
     {{-26\,\pi }\over {21}} + {{\gamma\,\pi }\over 2} + 
     {{67\,m\,q}\over {1440}} + {i\over {144}}\,m\,\pi \,q + 
     {i\over {360}}\,{q^2} - {{17\,i}\over {12960}}\,{m^2}\,{q^2} + 
     {{\pi \,\ln 2}\over 2}. 
\end{eqnarray}
Then noting that $\exp(-i\phi)\sim\exp(-i(z^*-z))$ at $z\sim\infty$, 
the asymptotic form of $X^{{\rm in}}_{\ell m\omega}$ is expressed as
\begin{eqnarray}
X^{{\rm in}}_{\ell m\omega}&=& \sqrt{z^2+a^2\omega^2}\exp(-i\phi)
\left\{f^{(0)}_{\ell m}+\epsilon\xi^{(1)}_{\ell m}
+\epsilon^2\xi^{(2)}_{\ell m}+...\right\}\nonumber\\
&\sim&e^{-i z^*}(z h_\ell^{(2)}e^{i z})\left[1
+\epsilon(p^{(1)}_{\ell m}+i q^{(1)}_{\ell m})
+\epsilon^2(p^{(2)}_{\ell m}+i q^{(2)}_{\ell m})\right]\nonumber\\
& &+e^{i z^*}(z h_\ell^{(1)}e^{-i z})\left[1
+\epsilon(p^{(1)}_{\ell m}-i q^{(1)}_{\ell m})
+\epsilon^2(p^{(2)}_{\ell m}-i q^{(2)}_{\ell m})\right],
\end{eqnarray}
where $h^{(1)}_{\ell}$ and $h^{(2)}_\ell$ are the spherical Hankel functions
of the first and second kinds, respectively, which are given by
\begin{equation}
h^{(1)}_{\ell}=j_\ell+i n_\ell \rightarrow (-1)^{\ell+1}{e^{i z}\over z},~ 
h^{(2)}_{\ell}=j_\ell-i n_\ell \rightarrow (-1)^{\ell+1}{e^{-i z}\over z}.
\end{equation} 
\ From these equations, noting $\omega r^*=z^*-\epsilon\ln\epsilon$, 
we obtain
\begin{equation}
A^{{\rm in}}_{\ell m\omega}={1\over 2}i^{\ell+1}e^{-i\epsilon\ln\epsilon}
\left[1+\epsilon(p^{(1)}_{\ell m}+i q^{(1)}_{\ell m})
+\epsilon^2(p^{(2)}_{\ell m}+i q^{(2)}_{\ell m})+...\right]. 
\end{equation}
The corresponding incident amplitude $B^{{\rm in}}_{\ell m\omega}$ 
for the Teukolsky function are obtained from Eq.(\ref{eq:ainbin}). 

\section{Gravitational wave luminosity to $O(v^8)$}
\subsection{The geodesic equations}
In this section,we solve the geodesic equation for circular motion in
the equatorial plane. 
The geodesic equations in the Kerr geometry are given by
\begin{eqnarray}
& &\Sigma{d\theta \over d\tau}=\pm\Bigl[C-\cos^2\theta \Bigl\{a^2(1-E^2)+
{l_z^2 \over \sin^2\theta}\Bigr\}\Bigr]^{1/2} \equiv \Theta(\theta),\nonumber\\
& &\Sigma{d\varphi \over d\tau}=-\Bigl(aE-{l_z \over \sin^2\theta}\Bigr)
+{a \over \Delta}\Bigl(E(r^2+a^2)-al_z\Bigr) \equiv \Phi, \nonumber\\
& &\Sigma{dt \over d\tau}=
-\Bigl(aE-{l_z \over \sin^2\theta}\Bigr)a\sin^2\theta
+{r^2+a^2 \over \Delta}\Bigl(E(r^2+a^2)-al_z\Bigr) \equiv T, \nonumber\\
& &\Sigma{dr \over d\tau}=\pm\sqrt{R},  \label{eq:geod}\\
\nonumber
\end{eqnarray}
where $E$, $l_z$ and $C$ are the energy, the $z$-component of the
angular momentum and the Carter constant of a test particle, respectively.
 $\Sigma=r^2+a^2\cos^2\theta$ and 
\begin{equation}
R=[E(r^2+a^2)-al_z]^2-\Delta[(Ea-l_z)^2+r^2+C].
\end{equation}
Since we consider a motion of a particle 
in the equatorial plane $\theta=\pi/2$, we can set $C=0$. 
We define the orbital radius as $r=r_0$. 
Then $E$ and $l_z$ are determined by $R(r_0)=0$ and 
$\partial R/\partial r\mid_{r=r_0}=0$ as
\begin{eqnarray*}
E&=&{{1-2v^2+q v^3}\over (1-3v^2+2qv^3)^{1/2}}\,,\\ 
l_z&=&{{r_0v(1-2qv^3+q^2 v^4)}\over 
(1-3v^2+2qv^3)^{1/2}}\,,\\ 
\end{eqnarray*}
where $v=(M/r_0)^{1/2}$. 
After these preparations, we can easily obtain $\varphi(t)$ as
\begin{eqnarray}
\varphi(t)&=&\Omega \, t\,, \nonumber\\
\Omega&=&{M^{1/2}\over r_0^{3/2}}
\left[1 - q v^3+q^2 v^6+O(v^9)\right]\,. 
\end{eqnarray} 

\subsection{Integration of the source term} 
Using results of the previous section, 
we can now derive the source term 
of the Teukolsky equation and integrate it to give the amplitude of the 
Teukolsky function at infinity. 

The energy momentum tensor of a test particle of mass $\mu$ is given by
\begin{equation}
T^{\mu\nu}={\mu\over \Sigma \sin\theta dt/d\tau}{dz^{\mu}\over d\tau}
{dz^{\nu}\over d\tau}\delta(r-r_0)\delta(\theta-\pi/2)
\delta(\varphi-\varphi(t)).
\end{equation}
The source term of the Teukolsky equation is given by 
\begin{equation}
T_{\ell m\omega}=4\int d\Omega dt\rho^{-5}{\bar\rho}^{-1}(B_2'+B_2'^*)
e^{-im\varphi+i\omega t}{_{-2}S^{a\omega}_{\ell m} \over
\sqrt{2\pi}},
\label{eq:source}
\end{equation}
where
\begin{eqnarray}
B_2'&=&-{1 \over 2}\rho^8{\bar \rho}L_{-1}[\rho^{-4}L_0
(\rho^{-2}{\bar \rho}^{-1}T_{nn})]  \nonumber\\
&&-{1 \over 2\sqrt{2}}\rho^8{\bar \rho}\Delta^2 L_{-1}[\rho^{-4}
{\bar \rho}^2 J_+(\rho^{-2}{\bar \rho}^{-2}\Delta^{-1}
T_{{\bar m}n})],  \nonumber\\
B_2'^*&=&-{1 \over 4}\rho^8{\bar \rho}\Delta^2 J_+[\rho^{-4}J_+
(\rho^{-2}{\bar \rho}T_{{\bar m}{\bar m}})] \nonumber\\
&&-{1 \over 2\sqrt{2}}\rho^8{\bar \rho}\Delta^2 J_+[\rho^{-4}
{\bar \rho}^2 \Delta^{-1} L_{-1}(\rho^{-2}{\bar \rho}^{-2}
T_{{\bar m}n})], \label{eq:B2}\\
\nonumber
\end{eqnarray}
with
\begin{eqnarray}
\rho&=&(r-ia\cos\theta)^{-1}, \nonumber\\
L_s&=&\partial_{\theta}+{m \over \sin\theta}
-a\omega\sin\theta+s\cot\theta, \label{eq:rho}\\
J_+&=&\partial_r+{iK/\Delta}, \nonumber\\
\nonumber
\end{eqnarray}
and ${\bar \rho}$ denotes the complex conjugate of $\rho$. 

In the present case, the tetrad components of the energy momentum
tensor, $T_{n\,n}$, $T_{{\bar m}\,n}$ and
$T_{{\bar m}\,{\bar m}}$, take the form,
\begin{eqnarray}
T_{n\,n}&=&{C_{n\,n} \over \sin\theta}
\delta(r-r_0) \delta(\theta-\pi/2) \delta(\varphi-\varphi(t)),\nonumber\\
T_{{\bar m}\,n}&=&{C_{{\bar m}\,n} \over \sin\theta}
\delta(r-r_0) \delta(\theta-\pi/2) \delta(\varphi-\varphi(t)),\label{eq:tnn}\\
T_{{\bar m}\,{\bar m}}&=&
{C_{{\bar m}\,{\bar m}} \over \sin\theta}
\delta(r-r_0) \delta(\theta-\pi/2) \delta(\varphi-
\varphi(t)),\nonumber\\
\nonumber
\end{eqnarray}
where
\begin{eqnarray}
C_{n\,n}&=&{\mu \over 4\Sigma^3 \dot t}\left[E(r^2+a^2)-al_z
\right]^2,\nonumber\\
C_{{\bar m}\,n}&=&
-{\mu \rho \over 2\sqrt{2}\Sigma^2 \dot t}\left[E(r^2+a^2)-al_z
\right]
\left[i\sin\theta\Bigl(aE-{l_z \over \sin^2\theta}\Bigr)\right], 
\label{eq:cnn}\\
C_{{\bar m}\,{\bar m}}&=&
{\mu \rho^2 \over 2\Sigma \dot t }
\left[i\sin\theta 
\Bigl(aE-{l_z \over \sin^2\theta}\Bigr)\right]^2,\nonumber\\
\nonumber
\end{eqnarray}
and $\dot t=dt/d\tau$.

Substituting Eq.(\ref{eq:B2}) into Eq.(\ref{eq:source}) 
and integrating by parts, we obtain
\begin{eqnarray}
T_{\ell m\omega}&=&{4 \over \sqrt{2\pi}}\int^{\infty}_{-\infty}
dt\int d\theta e^{i\omega t-im\varphi(t)} \nonumber\\
&\times&
\Bigl[-{1 \over 2}L_1^{\dag} \bigl\{ \rho^{-4}L_2^{\dag}(\rho^3 S)
\bigr\}
C_{n\,n}\rho^{-2}{\bar \rho}^{-1}\delta(r-r_0)
\delta(\theta-\pi/2) \nonumber\\
&+&{\Delta^2 {\bar \rho}^2 \over \sqrt{2} \rho}
\bigl(L_2^{\dag} S+ia({\bar \rho}-\rho)\sin\theta S\bigr)
J_+ \bigl\{ C_{{\bar m}\,n}\rho^{-2}{\bar \rho}^{-
2}\Delta^{-1}
\delta(r-r_0)\delta(\theta-\pi/2) \bigr\} \nonumber\\
&+&{1 \over 2\sqrt{2} }
L_2^{\dag}\bigl\{ \rho^3 S({\bar \rho}^2 \rho^{-4})_{,r} \bigr\}
C_{{\bar m}\,n}\Delta \rho^{-2}{\bar \rho}^{-2}
\delta(r-r_0)\delta(\theta-\pi/2) \nonumber\\
&-&{1 \over 4}\rho^3 \Delta^2 S J_+\bigl\{\rho^{-4}
J_+\bigl({\bar \rho} \rho^{-2}C_{{\bar m}\,{\bar m}}
\delta(r-r_0)\delta(\theta-\pi/2)\bigr) \bigr\}
\Bigr],\label{eq:tass}\\
\nonumber
\end{eqnarray}
where
\begin{equation}
L_s^{\dag}=\partial_{\theta}-{m \over \sin\theta}
+a\omega\sin\theta+s\cot\theta, 
\end{equation}
and $S$ denotes $_{-2}S_{\ell m}^{a\omega}(\theta)$ for simplicity.

We further rewrite Eq.(\ref{eq:tass}) as
\begin{eqnarray}
T_{\ell m\omega}&=&\int^{\infty}_{-\infty}dt e^{i\omega t-i m \varphi(t)}
\Delta^2\left[(A_{n\,n\,0}+A_{{\bar m}\,n\,0}+
A_{{\bar m}\,{\bar m}\,0})\delta(r-r_0) \right.\nonumber\\
&+&\left.
\left\{(A_{{\bar m}\,n\,1}+A_{{\bar m}\,{\bar m}\,1})
\delta(r-r_0)\right\}_{,r}
+\left\{A_{{\bar m}\,{\bar m}\,2}\delta(r-r_0)\right\}_{,rr}
\right]_{\theta=\pi/2}\,,
\label{eq:tone} \\
\nonumber
\end{eqnarray}
where $A_{n\,n\,0}$ etc. are given in Appendix B.
Inserting Eq.(\ref{eq:tone}) into Eq.(\ref{eq:Rinfty}), we obtain 
$\tilde Z_{\ell m\omega}$ as
\begin{eqnarray}
\tilde Z_{\ell m\omega}&=&
{{2 \pi \delta(\omega-m\Omega)}\over {2i\omega B^{{\rm in}}_{\ell m\omega}}}
\Bigl[R^{{\rm in}}_{\ell m\omega}\{A_{n\,n\,0}+A_{{\bar m}\,n\,0}
+A_{{\bar m}\,{\bar m}\,0}\}
\nonumber\\
& &-{dR^{{\rm in}}_{\ell m\omega} \over dr}\{ A_{{\bar m}\,n\,1}
+A_{{\bar m}\,{\bar m}\,1}\}
 +{d^2 R^{{\rm in}}_{\ell m\omega} \over dr^2}
 A_{{\bar m}\,{\bar m}\,2}\Bigr]_{r=r_0,\theta=\pi/2},
\nonumber\\
&\equiv& 
\delta(\omega-m\Omega) Z_{\ell m\omega}\,.
\label{eq:zzq}
\end{eqnarray} 
Using characters of $_{-2}S^{a\omega}_{\ell m}(\theta)$ 
at $\theta=\pi/2$, 
it is straightforward to show that $\bar{T}_{\ell, -m, -\omega}$ 
$=(-1)^{\ell}T_{\ell,m,\omega}$ 
where $\bar{T}_{\ell, m, \omega}$ is the complex conjugate of 
$T_{\ell,m,\omega}$. 
Since the homogeneous Teukolsky equation
is invariant 
under the complex conjugate followed by $m\rightarrow -m$ and
$\omega\rightarrow -\omega$, 
we have $\bar{Z}_{\ell, -m, -\omega}$ 
$=(-1)^{\ell}Z_{\ell,m,\omega}$. 

\subsection{Results}
In this section, we calculate the gravitational wave luminosity 
up to $O(v^8)$ beyond the quadrupole formula. From Eq.(\ref{eq:psifour}), 
$\psi_4$ at $r \rightarrow \infty$ takes a form,
\begin{equation}
\psi_4={1\over r}\sum_{\ell=2}^{6}\sum_{m=-\ell}^{\ell}
Z_{\ell m\omega_0}{{}_{-2}S_{\ell m}^{a\omega_0} \over \sqrt{2\pi}}
e^{i\omega_0(r^*-t)+im\varphi}\,, \label{eq:heq}
\end{equation}
where $\omega_0=m\Omega$. 
At infinity, $\psi_4$ is related to the two independent modes of
gravitational waves $h_+$ and $h_\times$ as
\begin{equation}
\psi_4={1\over2}(\ddot h_{+}-i\ddot h_{\times}).
\label{eq:hplcr}
\end{equation}
\ From Eqs.(\ref{eq:zzq}), (\ref{eq:heq}) and (\ref{eq:hplcr}), 
the gravitational wave luminosity is given by
\begin{equation}
\biggl< {dE \over dt} \biggr>=\sum_{\ell,m} 
{\Bigl|Z_{\ell m\omega_0}\Bigr|^2\over4\pi\omega_0^2} 
\equiv \sum_{\ell,m}\left( {dE \over dt} \right)_{\ell m}.\label{eq:flu}
\end{equation}
In order to express the post-Newtonian corrections to the luminosity,
we define $\eta_{\ell m}$ as 
\begin{equation}
\left( {dE \over dt} \right)_{\ell m}
\equiv{1\over2}\left({dE\over dt}\right)_N \eta_{\ell,m}\,,
\label{eq:etalmw}
\end{equation}
where $(dE/dt)_N$ is the Newtonian quadrupole luminosity:
$$
\left({dE\over dt}\right)_N={32\mu^2M^3\over5r_0^5}
={32\over5}\left({\mu\over M}\right)^2v^{10}.
\label{eq:Enewton}
$$
We only show $\eta_{\ell m}$
for $m> 0$ mode since $\eta_{\ell,m}=\eta_{\ell,-m}$. 

\begin{eqnarray}
\eta_{2,2}&=&
       1 - {{107\,{v^2}}\over {21}} + 
      \left( 4\,\pi  - 6\,q \right) \,{v^3} + 
      \left( {{4784}\over {1323}} + 2\,{q^2} \right) \,{v^4} 
\nonumber\\
&+& \left( {{-428\,\pi }\over {21}} + {{4216\,q}\over {189}} \right) \,
       {v^5} 
+ \left( {{99210071}\over {1091475}} - 
         {{1712\,\gamma }\over {105}} + {{16\,{{\pi }^2}}\over 3} 
\right.\nonumber\\
&-&\left.
 28\,\pi \,q + {{8830\,{q^2}}\over {567}} - 
     {{3424\,\ln 2}\over {105}} - {{1712\,\ln v}\over {105}} \right)
{v^6}
\nonumber\\
&+& \left( {{19136\,\pi }\over {1323}} + 
         {{163928\,q}\over {11907}} + 8\,\pi \,{q^2} - 12\,{q^3} \right) 
\,{v^7} 
\nonumber\\ 
&+&      \left( -{{27956920577}\over {81265275}} + 
         {{183184\,\gamma }\over {2205}} - {{1712\,{{\pi }^2}}\over {63}} + 
         {{20716\,\pi \,q}\over {189}} 
\right.
\nonumber\\
&-&\left. {{456028\,{q^2}}\over {9261}} + 
         {q^4} + {{366368\,\ln 2}\over {2205}} + 
         {{183184\,\ln v}\over {2205}} \right)\,{v^8}\,,
\\
\eta_{2,1}&=&
{{{v^2}}\over {36}} - {{q\,{v^3}}\over {12}} + 
      \left( -{{17}\over {504}} + {{{q^2}}\over {16}} \right) \,{v^4} 
\nonumber\\
&+&   \left( {{\pi }\over {18}} - {{793\,q}\over {9072}} \right) \,{v^5} + 
      \left( -{{2215}\over {254016}} - {{\pi \,q}\over 6} + 
         {{859\,{q^2}}\over {1512}} \right) \,{v^6} 
\nonumber\\
&+&   \left( {{-17\,\pi }\over {252}} + {{11861\,q}\over {190512}} + 
         {{\pi \,{q^2}}\over 8} - {{7\,{q^3}}\over {12}} \right) \,{v^7} 
\nonumber\\
&+&   \left( {{15707221}\over {26195400}} - 
         {{107\,\gamma }\over {945}} + {{{{\pi }^2}}\over {27}} - 
         {{1045\,\pi \,q}\over {4536}} 
\right.
\nonumber\\
&+&\left.
         {{118943\,{q^2}}\over {190512}} + 
         {{{q^4}}\over {16}} - {{107\,\ln 2}\over {945}} - 
         {{107\,\ln v}\over {945}} \right)\,{v^8}\,.
\end{eqnarray}
Putting together the above results, we obtain $(dE/dt)_\ell$ 
$\equiv\sum_m(dE/dt)_{\ell m}$ for $\ell=2$ as
\begin{eqnarray}
\left(dE\over dt\right)_{2}&=&
\left(dE\over dt\right)_N\left\{
1 - {{1277\,{v^2}}\over {252}} + 
      \left( 4\,\pi  - {{73\,q}\over {12}} \right) \,{v^3} + 
      \left( {{37915}\over {10584}} + {{33\,{q^2}}\over {16}} \right) \,
       {v^4} 
\right.
\nonumber\\
&+& \left( {{-2561\,\pi }\over {126}} + 
         {{201575\,q}\over {9072}} \right) \,{v^5} 
  + \left( {{2116278473}\over {23284800}} - 
         {{1712\,\gamma }\over {105}} + {{16\,{{\pi }^2}}\over 3} 
\right. 
\nonumber\\
&-&\left. 
         {{169\,\pi \,q}\over 6} + {{73217\,{q^2}}\over {4536}} - 
         {{3424\,\ln 2}\over {105}} - {{1712\,\ln v}\over {105}}
 \right)\,{v^6}
\nonumber\\
&+& \left( {{76187\,\pi }\over {5292}} + {{376387\,q}\over {27216}} + 
         {{65\,\pi \,{q^2}}\over 8} - {{151\,{q^3}}\over {12}} \right) \,{v^7}
\nonumber\\
&+&   \left( -{{2455920939443}\over {7151344200}} + 
         {{548803\,\gamma }\over {6615}} - {{5129\,{{\pi }^2}}\over {189}} + 
         {{70877\,\pi \,q}\over {648}} 
\right.
\nonumber\\
&-& \left.\left.
{{64835431\,{q^2}}\over {1333584}} + 
         {{17\,{q^4}}\over {16}} + {{219671\,\ln 2}\over {1323}} + 
         {{548803\,\ln v}\over {6615}} \right)\,{v^8}\right\}\,.
\end{eqnarray}
For $\ell=3$, we obtain
\begin{eqnarray}
\eta_{3,3}&=&
 {{1215\,{v^2}}\over {896}} - {{1215\,{v^4}}\over {112}} + 
      \left( {{3645\,\pi }\over {448}} - {{1215\,q}\over {112}} \right) \,
       {v^5} 
\nonumber\\
&+& \left( {{243729}\over {9856}} + {{3645\,{q^2}}\over {896}}
          \right) \,{v^6} + \left( {{-3645\,\pi }\over {56}} + 
         {{131949\,q}\over {1792}} \right) \,{v^7} 
\nonumber\\
&+& 
   \left( {{25037019729}\over {125565440}} - 
         {{47385\,\gamma }\over {1568}} + {{3645\,{{\pi }^2}}\over {224}} - 
         {{32805\,\pi \,q}\over {448}} 
\right.
\nonumber\\
&+&\left.
 {{346275\,{q^2}}\over {14336}} - 
         {{47385\,\ln 2}\over {1568}} - {{47385\,\ln 3}\over {1568}} - 
         {{47385\,\ln v}\over {1568}} \right) \,v^8 \,,
\\
\eta_{3,2}&=&
{{5\,{v^4}}\over {63}} - {{40\,q\,{v^5}}\over {189}} + 
      \left( -{{193}\over {567}} + {{80\,{q^2}}\over {567}} \right) \,{v^6} 
\nonumber\\
&+& 
      \left( {{20\,\pi }\over {63}} + {{352\,q}\over {1701}} \right) \,
       {v^7} + \left( {{86111}\over {280665}} - {{160\,\pi \,q}\over {189}} + 
         {{40\,{q^2}}\over {27}} \right) \,{v^8} \,,
\\
\eta_{3,1}&=&{{{v^2}}\over {8064}} - {{{v^4}}\over {1512}} + 
      \left( {{\pi }\over {4032}} - {{17\,q}\over {9072}} \right) \,{v^5} 
\nonumber\\
&+& 
 \left( {{437}\over {266112}} + {{17\,{q^2}}\over {24192}} \right) \,
       {v^6} + \left( {{-\pi }\over {756}} + {{3601\,q}\over {435456}} \right)
        \,{v^7} 
\nonumber\\
&+& \left( -{{1137077}\over {50854003200}} - 
         {{13\,\gamma }\over {42336}} + {{{{\pi }^2}}\over {6048}} - 
         {{145\,\pi \,q}\over {36288}} + {{41183\,{q^2}}\over {3483648}} 
\right.
\nonumber\\
&-&\left.
 {{13\,\ln 2}\over {42336}} - {{13\,\ln v}\over {42336}} \right)
{v^8} \,.
\end{eqnarray}
Then we obtain
\begin{eqnarray}
\left(dE\over dt\right)_{3}&=&
\left(dE\over dt\right)_N\left\{
 {{1367\,{v^2}}\over {1008}} - {{32567\,{v^4}}\over {3024}} + 
      \left( {{16403\,\pi }\over {2016}} - {{896\,q}\over {81}} \right) \,
       {v^5} 
\right.
\nonumber\\
&+& 
\left( {{152122}\over {6237}} + {{341\,{q^2}}\over {81}}
          \right) \,{v^6} + \left( {{-13991\,\pi }\over {216}} + 
         {{4019665\,q}\over {54432}} \right) \,{v^7} 
\nonumber\\
&+& 
 \left( {{5712521850527}\over {28605376800}} - 
         {{79963\,\gamma }\over {2646}} + {{6151\,{{\pi }^2}}\over {378}} - 
         {{192005\,\pi \,q}\over {2592}} 
\right. 
\nonumber\\
&+& \left. \left.
         {{11168371\,{q^2}}\over {435456}} - 
         {{79963\,\ln 2}\over {2646}} - {{47385\,\ln 3}\over {1568}} - 
         {{79963\,\ln v}\over {2646}} \right)\,v^8\right\} \,.
\end{eqnarray}
For $\ell=4$, we have 
\begin{eqnarray}
\eta_{4,4}&=&
{{1280\,{v^4}}\over {567}} - {{151808\,{v^6}}\over {6237}} + 
      \left( {{10240\,\pi }\over {567}} - {{12800\,q}\over {567}} \right) \,
       {v^7} 
\nonumber\\
&+& \left( {{560069632}\over {6243237}} + 
         {{5120\,{q^2}}\over {567}} \right) \,{v^8}\,,
\\
\eta_{4,3}&=&
{{729\,{v^6}}\over {4480}} - {{729\,q\,{v^7}}\over {1792}} + 
      \left( -{{28431}\over {24640}} + {{3645\,{q^2}}\over {14336}} \right) \,
       {v^8}\,,
\\
\eta_{4,2}&=& {{5\,{v^4}}\over {3969}} - {{437\,{v^6}}\over {43659}} + 
      \left( {{20\,\pi }\over {3969}} - {{80\,q}\over {3969}} \right) \,
       {v^7} 
\nonumber\\
&+& \left( {{7199152}\over {218513295}} + 
         {{200\,{q^2}}\over {27783}} \right) \,{v^8}\,,
\\
\eta_{4,1}&=& {{{v^6}}\over {282240}} - {{q\,{v^7}}\over {112896}} + 
      \left( -{{101}\over {4656960}} + {{5\,{q^2}}\over {903168}} \right) \,
       {v^8}\,.
\end{eqnarray}
Then we obtain
\begin{eqnarray}
\left(dE\over dt\right)_{4}&=&
\left(dE\over dt\right)_N\left\{
 {{8965\,{v^4}}\over {3969}} - 
      {{84479081\,{v^6}}\over {3492720}} + 
      \left( {{23900\,\pi }\over {1323}} - {{59621\,q}\over {2592}} \right) \,
       {v^7} 
\right.
\nonumber\\
&+&\left.
 \left( {{51619996697}\over {582702120}} + 
         {{66084895\,{q^2}}\over {7112448}} \right) \,{v^8}\right\}\,.
\end{eqnarray}
For $\ell=5$ we have 
\begin{eqnarray}
\eta_{5,5}&=&
 {{9765625\,{v^6}}\over {2433024}} - 
      {{2568359375\,{v^8}}\over {47443968}}\,,
\\
\eta_{5,4}&=&{{4096\,{v^8}}\over {13365}}\,,
\\
\eta_{5,3}&=& {{2187\,{v^6}}\over {450560}} - 
      {{150903\,{v^8}}\over {2928640}}\,,
\\
\eta_{5,2}&=&{{4\,{v^8}}\over {40095}}\,,
\\
\eta_{5,1}&=&{{{v^6}}\over {127733760}} - 
      {{179\,{v^8}}\over {2490808320}}\,.
\end{eqnarray}
Then we have
\begin{eqnarray}
\left(dE\over dt\right)_{5}&=&
\left(dE\over dt\right)_N\left\{
{{1002569\,{v^6}}\over {249480}} - 
      {{3145396841\,{v^8}}\over {58378320}}\right\}\,.
\end{eqnarray}
For $\ell=6$ we have 
\begin{eqnarray}
\eta_{6,6}&=&
{{26244\,{v^8}}\over {3575}}\,,
\\
\eta_{6,4}&=&{{131072\,{v^8}}\over {9555975}}\,,
\\
\eta_{6,2}&=&{{4\,{v^8}}\over {5733585}}\,,
\end{eqnarray}
and $\eta_{6,5}$, $\eta_{6,3}$, $\eta_{6,1}$ become $O(v^9)$. 
Then we have
\begin{eqnarray}
\left(dE\over dt\right)_{6}&=&
\left(dE\over dt\right)_N\,
{{210843872\,{v^8}}\over {28667925}}\,.
\end{eqnarray}

Finally, gathering all the above results, the total luminosity 
up to $O(v^8)$ is expressed as
\begin{eqnarray}
\biggl<{dE\over dt}\biggr>&=&\left({dE\over dt}\right)_N
\left\{1 - {{1247\,{v^2}}\over {336}} + 
      \left( 4\,\pi  - {{73\,q}\over {12}} \right) \,{v^3} + 
      \left( -{{44711}\over {9072}} + {{33\,{q^2}}\over {16}} \right) \,
       {v^4} 
\right.
\nonumber\\
&+& \left( {{-8191\,\pi }\over {672}} + {{3749\,q}\over {336}}
          \right) \,{v^5} 
+ \left( {{6643739519}\over {69854400}} - 
         {{1712\,\gamma }\over {105}} 
\right.
\nonumber\\
&+& \left.
    {{16\,{{\pi }^2}}\over 3} - 
    {{169\,\pi \,q}\over 6} + {{3419\,{q^2}}\over {168}} - 
    {{3424\,\ln 2}\over {105}} - {{1712\,\ln v}\over {105}} \right)
\,{v^6}
\nonumber\\
&+& \left( {{-16285\,\pi }\over {504}} + 
         {{83819\,q}\over {1296}} + {{65\,\pi \,{q^2}}\over 8} - 
         {{151\,{q^3}}\over {12}} \right) \,{v^7} 
\nonumber\\
&+& \left( -{{323105549467}\over {3178375200}} + 
         {{232597\,\gamma }\over {4410}} - {{1369\,{{\pi }^2}}\over {126}} + 
         {{3389\,\pi \,q}\over {96}} - {{124091\,{q^2}}\over {9072}} 
\right.
\nonumber\\
&+&\left.\left.
     {{17\,{q^4}}\over {16}} + {{39931\,\ln 2}\over {294}} - 
     {{47385\,\ln 3}\over {1568}} + {{232597\,\ln v}\over {4410}}
          \right)\,{v^8}\right\}\,.
\label{eq:dedtall}
\end{eqnarray}
In Appendix G, we present formulas for $\eta_{\ell,m}$ and $dE/dt$
in terms of $v'\equiv(M\Omega)^{1/3}$ for the sake of 
convenience to calculate the phase function for a inspiraling wave 
form\cite{Ref:flanagan}. 

Setting $q=0$, above reproduces the previous 
results \cite{Ref:TN}\cite{Ref:TS} in a Schwarzschild case. 
Up to $O(v^5)$, the results 
agree  with those obtained by SSTT\cite{Ref:SSTT} 
in the case when the test particle moves a circular orbit in the 
equatorial plane. 
For $\ell=5$ and $6$, there are no contributions due to the black hole 
spin and the results are identical to the Schwarzschild case. 

In Eq.(\ref{eq:dedtall}), the numerical value of terms at order
$O(v^6)$ is given by
$v^6(115.7 - 88.48 q $ $+ 20.35 q^2 - 16.30 \ln v)$. 
We find that the spin dependent terms are not so small compared to the 
other two terms if $|q|$ is of order unity.
Thus, we see that spin dependent terms at $O(v^6)$ will 
give a significant effect to template wave forms of coalescing 
binaries when spin of a black hole is large. 

Finally we note that the angular momentum flux can be easily calculated from 
\begin{eqnarray}
\biggl<{dJ \over dt}\biggr>={1 \over \Omega}\biggl<{dE \over dt}\biggr>. 
\label{eq:dldtall} 
\end{eqnarray} 

\section{Comparison with numerical results} 

As discussed in section I, it is important to investigate 
the detailed convergence property of the post-Newtonian approximation. 
Therefore we compare the formula for  
$dE/dt$, derived above, with numerical results and investigate 
the accuracy of the post-Newtonian expansion of $dE/dt$. 

In this section, we consider the total mass of the 
binary systems including 
black holes  $\sim 2 - 300M_{\odot}$ 
because gravitational waves from such binaries 
can be detected by LIGO and VIRGO. 
In particular, we pay attention to 
the accuracy of post-Newtonian formula for 
$dE/dt$ when $r \leq 100M$ (or $v \geq 0.1 $), 
because gravitational waves from these binary systems 
will be detected 
when the orbital separation becomes less than $r \simeq 100M$. 
Here, we ignore the effect of absorption of gravitational waves by the  
black hole. 
We will briefly discuss its effect in the next section. 

A numerical study of $dE/dt$ 
from a particle in a circular orbit in the equatorial plane 
around a Kerr black hole has been performed by Shibata\cite{Ref:shibata}. 
Since nothing was assumed about the velocity of
a test particle, 
those results are correct relativistically 
in the limit $\mu\ll M$. 
In that work, $dE/dt$ was calculated with accuracy 
$\lesssim 10^{-4}$. However we found that this accuracy is not sufficient 
to compare it with the post-Newtonian formula for $dE/dt$ 
including terms up to $O(v^8)$. 
Thus, in this paper, we calculate $dE/dt$ again 
requiring the accuracy to be $\sim 10^{-5}$. 
In the numerical calculations, we have taken into 
account the contribution from the 
$\ell=2$ through $\ell=6$ 
modes in $dE/dt$ which is consistent with the post-Newtonian
formula. 

In figs.1(a$-$e), we show the error in the post-Newtonian formulas as 
a function of the Boyer-Lindquist coordinate radius 
when $q=-0.9$, $-0.5$, $0$, 0.5 and $0.9$. 
In these figures, we show the error for $6 \leq r/M \leq 100$. 
Since the radius of the inner stable circular orbit for $q=0.9$ is 
$r_{lso}\simeq 2.32M$ and a stable circular orbit is possible for $r >
r_{lso}$, we also show the errors in the case when $q=0.9$ for 
$2.5 \leq r/M \leq 12$ in fig.2. 
The error in the post-Newtonian formula is defined as 
\begin{eqnarray}
{\rm Error}=\left|1-\left({dE \over dt}\right)_{\rm PN}
\bigg/ \left({dE \over dt}\right)_{\rm NR}\right|, 
\end{eqnarray} 
where $(dE/dt)_{\rm PN}$ and $(dE/dt)_{\rm NR}$ denote the 
post-Newtonian formula and the numerical results respectively. 
As for $(dE/dt)_{\rm PN}$, 
we have used 2-PN, 2.5-PN, 3-PN, 3.5-PN and 4-PN formulas. 
Here, we define n-PN formula as the expression for 
$dE/dt$ which includes post-Newtonian terms up to $O(v^{2n})$ beyond 
the quadrupole formula. 
In each figure, open square, filled triangle, open triangle, 
filled circle, and open circle denote the error of 
2-PN, 2.5-PN, 3-PN, 3.5-PN and 4-PN formulas, respectively. 
We note that in fig.2, the errors in the 2.5-, 3- and 4-PN formulas 
become greater than unity for very small radius, because 
in such a region, $dE/dt$ for those PN formulas becomes negative. 

\ From these figures, we find the following. 

\noindent
(1) If we use the 2-PN or 2.5-PN formula, the error 
is always greater than $10^{-4}$ when $r \lesssim 100M$ 
irrespective of $q$. 
If we use the 3-PN formula, however, the error decreases significantly, and 
it becomes less than $10^{-4}$ for $r > 60M$, and less than $10^{-3}$ 
for $r > 30M$ irrespective of $q$. 

\noindent 
(2) If we adopt the 3.5-PN formula, the accuracy becomes better than
that of 3-PN formula. 
The error is always less than $10^{-4}$ when $r$ is greater than 
$\sim 30M$ and less than 
$10^{-5}$ when $r$ is greater than $\sim 60M$. 
This feature does not depend on $q$. 
However, if we use the 4-PN formula, the accuracy is not 
improved compared with the 3.5-PN formula. In particular, this
tendency is remarkable for smaller radius. 

\noindent
(3) The accuracy of the 3.5-PN or 4-PN 
formula is not always better than that of the lower-PN 
one inside $r_c$, where $r_c \lesssim 5M$ for $q=0.5$ and $0.9$, 
$r_c \sim 10M$ for $q=0$ and $-0.5$, and $r_c \sim 15M$ for $q=-0.9$. Thus, 
the convergence of the post-Newtonian expansion seems rather poor around 
$r_c$. 

Using the above results, we investigate the accuracy of the 
post-Newtonian formulas as templates for various binary systems. 
As explained in section 1, to investigate the accuracy of 
the post-Newtonian formulas as templates, 
it is useful to check if 
they can predict the number of 
cycles of the gravitational waves, $N$, with accuracy less than 1. 
compact binary systems, the cycles  
are mainly accumulated around $\sim 10$Hz which is the 
lowest frequency region in the LIGO band, and $N$ is 
approximately given by 
\begin{eqnarray} 
N \sim 1.9 \times 10^3 \left({10M_{\odot} \over M} \right)^{5/3}
\left({M \over 4\mu} \right), 
\end{eqnarray} 
where $M$ and $\mu$ are the total mass and reduced mass, respectively. 
This means that the template must have an accuracy less than 
\begin{eqnarray}
\sim 5\times10^{-4} \left({M \over 10M_{\odot}} \right)^{5/3}
\left({4\mu \over M} \right), 
\end{eqnarray} 
when the frequency of gravitational wave becomes $10$Hz. 

First we consider equal mass binary systems, that is $M=4\mu$. 
At $10$Hz, the orbital separation of a binary of total mass $M$ 
is approximately given by $r/M \simeq 347(M_{\odot}/M)^{2/3}$. 
We find that 
the 2-PN and 2.5-PN formulas are insufficient if $M \gtrsim 
5M_{\odot}$, 
and  the 3-PN formula is needed. 
The 3-PN formula seems adequate irrespective of $q$. 

On the other hand, the 
situation is slightly different in the case when 
a neutron star of mass $\sim 1.4M_{\odot}$ 
spirals into a larger black hole. 
In such a case, the number of the 
cycles of the gravitational waves is large compared with
the equal mass case when the total mass is the  same. 
Thus, it seems that we need at least the 3.5-PN formula 
for binaries of mass greater than $\sim 30M_{\odot}$ 
to obtain the required accuracy. 
Also, for binaries of mass greater than $\sim 70M_{\odot}$, 
we need higher post-Newtonian corrections beyond 4-PN order. 

Binary systems of total mass greater than $\sim 100M_{\odot}$
can be detected when $r$ is smaller than $\sim 15M$. 
However, 
as mentioned in (3) above, the convergence property of the post-Newtonian 
expansion becomes bad for small orbital separations. 
In particular, for $q \sim -1$, the accuracy of the 
post-Newtonian expansion seems bad at $r \sim 15M$. 
Thus, it may not be appropriate to 
use the post-Newtonian approximation for binaries 
of total mass $\sim 100M_{\odot}$ with large mass ratio $\mu\ll M$. 
A more detailed investigations of the 
convergence of the post-Newtonian expansion will require the 
calculation to be carried beyond 4-PN order. 

\section{Summary and Discussion}

In this paper, we have performed a post-Newtonian expansion of gravitational
waves from a particle in a circular orbit around
a Kerr black hole. The orbit lies in the equatorial 
plane and the calculations are accurate to 
$O(v^8)$ beyond the quadrupole level. 
We have performed the post-Newtonian expansion of the 
Sasaki-Nakamura equation 
and obtained the Green function of the radial Teukolsky equation 
up to $O(\epsilon^2)$ using methods developed previously. 
Then we obtained all the necessary radial 
functions to the required accuracy. We have also calculated the spin weighted 
spheroidal harmonics up to $O((a\omega)^2)$. The outgoing wave 
amplitude of the Teukolsky function and the  
gravitational wave luminosities were 
derived up to $O(v^8)$ beyond the quadrupole formula. 

It is worth noting that 
in the formula for $\eta_{2,2}$ in Appendix G, there are terms such as 
$(-8/3) q v'^3$, $2 q^2 v'^4$, $(-8/3) q^3 v'^7$ and $q^4 v'^8$. 
In a previous paper\cite{Ref:SSTT}, 
we pointed out that the term $2 q^2 v'^4$ 
can be explained in terms of the quadrupole formula as the contribution
of the quadrupole moment of the Kerr black hole to the orbit of the test
particle. 
A similar explanation is possible for $(-8/3) q^3 v'^7$ 
and $q^4 v'^8$. 
We can derive those terms by using the quadrupole formula 
$dE/dt=32/5 \mu^2 \hat r^4\Omega^6$, where $\hat r$ is the orbital radius of
a test particle in de Donder coordinates. 
If multipole moments of the black hole exist, the orbital radius is changed 
due to the influence of those multipole moments 
(or if we fix the orbital radius, $\Omega$ 
is changed due to the multipole moments of black hole). 
We can calculate the leading order effect of the multipole moments 
to the orbital radius by using multipole expansion of 
the Kerr metric (Eq.(10.6) of Ref.\cite{Ref:kip}). 
In this way,  we find 
that the dominant effect of the multipole moments of a Kerr 
black hole to $dE/dt$ can be expressed as \cite{Ref:fintan1}
\begin{equation} 
{dE\over dt}={32\over 5}\left(\mu\over M \right)^2v'^{10}\left\{
1-{8\over 3}S_1 v'^3-2 M_2 v'^4+4 S_3 v'^7 +
\left(-{3\over 2}M_2^2+{5\over 2}M_4\right)v'^8\right\}, 
\end{equation} 
where $M_\ell$ and $S_\ell$ are mass and current multipole moments  
of a Kerr black hole given by 
$M_\ell+iS_\ell=M(ia)^\ell$. 
Now we can interpret 
the term $-12 q^3 v^7$ as the effect of 
the current octopole moment of a black hole and 
the term $q^4 v^8$ as the effect of both the 
mass quadrupole moment and $\ell=4$ mass multipole moment of 
a black hole. 

As for $\ell=2$ and $m=1$ mode, there are terms $-q v'^3/12$, $q^2 v'^4/16$, 
$-7q^3 v'^7/24$ and $q^4 v'^8/16$. The terms $-q v'^3/12$ and 
$q^2 v'^4/16$ can be explained as the correction to the 
radiative current quadrupole moment \cite{Ref:KWW}~\cite{Ref:fintan2}. 
We expect that the terms $-7q^3 v'^7/24$ and $q^4 v'^8/16$  
can also be derived simply in a similar way. 

In section 4, 
by comparing post-Newtonian formulas for $dE/dt$ with 
numerical data, we indicated that the 
convergence of the post-Newtonian expansion 
seems bad when orbital radii of binaries become less than $\sim 15M$. 
This suggests that the post-Newtonian expansion may not be 
appropriate to construct theoretical templates 
for large mass ratio 
binaries where  the total mass is greater than 
$\sim 100M_\odot$ 
because gravitational waves from such 
binaries enter the LIGO/VIRGO 
frequency band when $r \lesssim 15M$. 
Nevertheless, the higher order post-Newtonian terms 
gradually improve the accuracy of the templates. Hence, it is 
very natural to ask whether the post-Newtonian expansion is always 
appropriate or not, and if appropriate, up to what order we need the
post-Newtonian terms to construct accurate templates. 
Fortunately, it is possible to obtain the formulas for $dE/dt$ 
which include post-Newtonian order terms beyond $O(v^8)$ 
by extending techniques developed in this paper. 
Extension of the present work up to the higher 
post-Newtonian order, beyond $O(v^8)$, is very important and 
that is our future work. 

The analysis, 
in this paper, has been restricted to the case when a test particle moves
in a circular orbit on the equatorial plane. However, as shown in a previous 
paper\cite{Ref:SSTT}, inclination of the orbital plane 
from the equatorial plane will significantly affect 
the orbital phase evolution. 
Hence, the present work should be considered as a first step toward the 
complete calculation of the energy and angular momentum luminosities including 
the orbital inclination. 

Finally, we comment on the effect of absorption of gravitational waves 
by the black hole event horizon which should be taken into account when we 
consider the orbital evolution of black hole binaries. According to 
Gal'tsov\cite{Ref:galt}, the lowest order contribution of the 
gravitational wave absorption to $dE/dt$ is given by  
\begin{eqnarray}
{dE \over dt}=\biggl( {dE \over dt} \biggr)_N {v^5 \over 2}
\left\{v^3 \Bigl(1+\sqrt{1-q^2}\Bigr)-{q \over 2} \right\}\left(1+3q^2\right). 
\end{eqnarray}
Thus, the effect of absorption appears from $O(v^5)$ if $q \not=0$. 
Although the coefficient is small compared with that of $dE/dt$ for 
the outgoing wave even in the case $|q| \sim 1$, we need the expression 
for $dE/dt$ due to the black hole absorption 
to obtain an accurate template up to $O(v^8)$. 
Therefore, to obtain 
the higher order post-Newtonian corrections to the black hole absorption 
is a problem for the future. 

\begin{center} 
\Large{\bf Acknowledgments} 
\end{center} 
H.T. thanks F. Ryan for useful comments and Kip Thorne for 
continuous encouragement. 
We thank P.R. Brady for careful reading of the manuscript. 
H.T. was supported by Research Fellowship  
of the Japan Society for the Promotion of Science for Young Scientists 
and by NSF Grant No. AST-9417371 and NASA Grand No. NAGW-4268. 
This work was also supported in part by the Japanese Grant-in-Aid for 
Scientific Research of the Ministry of Education, 
Science and Culture, No. 07740355 and No. 04234104.

\appendix
\section{The formulae of $F$ and $U$}

In this Appendix we show the potential functions $F$ and $U$ 
of the SN equation (\ref{eq:sneq}). Details of the derivation are 
given in Ref. \cite{Ref:SN}.

The function $F(r)$ is given by
\begin{equation}
F(r)={\eta_{,r} \over \eta}{\Delta \over r^2+a^2},
\label{eq:Fdef}
\end{equation}
where 
\begin{equation}
\eta=c_0+c_1/r+c_2/r^2+c_3/r^3+c_4/r^4,
\label{eq:etadef}
\end{equation}
with
\begin{eqnarray}
&c_0&=-12i\omega M+\lambda(\lambda+2)-12a \omega(a\omega-m),\nonumber\\
&c_1&=8ia[3a\omega-\lambda(a\omega-m)],\nonumber\\
&c_2&=-24ia M(a\omega-m)+12a^2[1-2(a\omega-m)^2 ],\nonumber\\
&c_3&=24ia^3(a\omega-m)-24Ma^2,\nonumber\\
&c_4&=12a^4.\label{eq:ceq}\\
\nonumber
\end{eqnarray}
The function $U(r)$ is given by
\begin{equation}
U(r)={\Delta U_1\over(r^2+a^2)^2}+G^2+{\Delta G_{,r}\over r^2+a^2}-FG,
\label{eq:Udef}
\end{equation}
where
\begin{eqnarray}
G&=&-{2(r-M) \over r^2+a^2}+{r\Delta \over (r^2+a^2)^2},\nonumber\\
U_1&=&V+{\Delta^2 \over \beta}
\Bigl[\Bigl(2\alpha+{\beta_{,r}  \over \Delta}\Bigr)_{,r}
-{\eta_{,r} \over \eta}
\Bigl(\alpha +{\beta_{,r} \over \Delta}\Bigr)\Bigr],\nonumber\\
\alpha&=&-i{K \beta \over \Delta^2}+3iK_{,r}
+\lambda+{6\Delta \over r^2},\nonumber\\
\beta&=&2\Delta\Bigl(-iK+r-M-{2\Delta \over r}\Bigr).\label{eq:ur}\\
\nonumber
\end{eqnarray}

\section{Functions in the source term}

In this Appendix, we show the $A$'s in Eq.(\ref{eq:source}).
\begin{eqnarray*}
A_{n\,n\,0}&=&{-2 \over \sqrt{2\pi}\Delta^2}
C_{n\,n}\rho^{-2}{\bar \rho}^{-1}
L_1^+\{\rho^{-4}L_2^+(\rho^3 S)\},\\
A_{{\bar m}\,n\,0}&=&{2 \over \sqrt{\pi}\Delta} 
C_{{\bar m}\,n}\rho^{-3}
\Bigl[\left(L_2^+S\right)
\Bigl({iK \over \Delta}+\rho+{\bar \rho}\Bigr)
\\
& &\qquad-a\sin\theta S {K \over \Delta}({\bar \rho}-\rho)\Bigr],\\
A_{{\bar m}\,{\bar m}\,0}
&=&-{1 \over \sqrt{2\pi}}\rho^{-3}{\bar \rho}
C_{{\bar m}\,{\bar m}}S\Bigl[
-i\Bigl({K \over \Delta}\Bigr)_{,r}-{K^2 \over \Delta^2}+
2i\rho {K \over \Delta}\Bigr],\\
A_{{\bar m}\,n\,1}&=&{2\over \sqrt{\pi}\Delta }
\rho^{-3}C_{{\bar m}\,n}
[L_2^+S+ia\sin\theta({\bar \rho}-\rho)S],\\
A_{{\bar m}\,{\bar m}\,1}
&=&-{2 \over \sqrt{2\pi}}
\rho^{-3}{\bar \rho}
C_{{\bar m}\,{\bar m}}S\Bigl(i{K \over \Delta}+\rho\Bigr),\\
A_{{\bar m}\,{\bar m}\,2}
&=&-{1\over \sqrt{2\pi}}\rho^{-3}{\bar \rho}
C_{{\bar m}\,{\bar m}}S,\\
\end{eqnarray*}
where $S$ denotes $_{-2}S_{\ell m}^{a\omega}$.

\section{$Q^{(2)},Q^{(3)},Q^{(4)}$}

\begin{eqnarray*}
Q^{(2)}&=&\left[ \left(-28\,i\,m\,q - 
         {{32\,i\,m\,q}\over {\ell}} + 
         8\,i\,\ell\,m\,q + 
         4\,i\,{{\ell}^2}\,m\,q - 
         13\,{q^2} - 
         {{6\,{q^2}}\over {\ell}} - 
         12\,\ell\,{q^2} - 
         {{\ell}^2}\,{q^2} + \right.\right. \\
& &\left.    6\,{{\ell}^3}\,{q^2} + 
         2\,{{\ell}^4}\,{q^2} + 
         8\,{m^2}\,{q^2} + 
         {{32\,{m^2}\,{q^2}}\over 
           {{{\ell}^2}}} + 
         {{8\,{m^2}\,{q^2}}\over {\ell}} \right)
         {1\over {{z^4}}}  + \\
& & \left(16\,m\,q + {{24\,m\,q}\over 
           {{{\ell}^2}}} + 
         {{20\,m\,q}\over {\ell}} - 
         8\,\ell\,m\,q - 
         4\,{{\ell}^2}\,m\,q - 
         14\,i\,{q^2} - 
         {{16\,i\,{q^2}}\over {\ell}} + \right.\\
& &   4\,i\,\ell\,{q^2} + 
         2\,i\,{{\ell}^2}\,{q^2} + 
         2\,i\,\lambda_1\,m\,{q^2} - 
         {{4\,i\,\lambda_1\,m\,{q^2}}\over 
           {{{\ell}^2}}} + \\
& &\left.   {{2\,i\,\lambda_1\,m\,{q^2}}\over 
           {\ell}} - 4\,i\,{m^2}\,{q^2} + 
         {{56\,i\,{m^2}\,{q^2}}\over 
           {{{\ell}^2}}} - 
         {{4\,i\,{m^2}\,{q^2}}\over {\ell}} \right)
        {1\over {{z^3}}} + \\
& &\left({{24\,i\,m\,q}\over {{{\ell}^2}}} + 
         {{17\,{q^2}}\over 2} + 
         {{10\,{q^2}}\over {\ell}} - 
         {{13\,\ell\,{q^2}}\over 4} - 
         {{9\,{{\ell}^2}\,{q^2}}\over 4} - 
         {{3\,{{\ell}^3}\,{q^2}}\over 4} - 
         {{{{\ell}^4}\,{q^2}}\over 4} - 
         {{\lambda_2\,{q^2}}\over 2} - \right.\\
& &   {{3\,\ell\,\lambda_2\,{q^2}}\over 
           4} + {{{{\ell}^2}\,\lambda_2\,
             {q^2}}\over 4} + 
         {{3\,{{\ell}^3}\,\lambda_2\,
             {q^2}}\over 4} + 
         {{{{\ell}^4}\,\lambda_2\,
             {q^2}}\over 4} - 
         2\,\lambda_1\,m\,{q^2} + \\
& &\left.\left.
         {{4\,\lambda_1\,m\,{q^2}}\over 
           {{{\ell}^2}}} - 
         {{2\,\lambda_1\,m\,{q^2}}\over 
           {\ell}} - 
         {{24\,{m^2}\,{q^2}}\over 
           {{{\ell}^2}}} \right)
      {1\over {{z^2}}}
      \right]{1\over {(\ell+1)^2(\ell^2+\ell-2)}}  + \\
& &  \left[ \left(-24\,i\,\lambda_0\,m\,q - 
         4\,i\,{{\lambda_0}^2}\,m\,q + 
         4\,i\,{{\lambda_0}^3}\,m\,q - 
         12\,\lambda_0\,{q^2} - 
         6\,{{\lambda_0}^2}\,{q^2} + \right.\right.\\
& &\left.      24\,\lambda_0\,{m^2}\,{q^2} - 
         4\,{{\lambda_0}^2}\,{m^2}\,{q^2}\right){1\over z^3} 
+ \left(24\,\lambda_0\,m\,q - 
         12\,i\,\lambda_0\,{q^2} - 
         2\,i\,{{\lambda_0}^2}\,{q^2} + 
         2\,i\,{{\lambda_0}^3}\,{q^2} + \right.\\
& &\left.\left. 2\,i\,{{\lambda_0}^2}\,\lambda_1\,m\,
          {q^2} + 24\,i\,\lambda_0\,{m^2}\,{q^2}\right)
       {1\over z^2} 
      \right]{1\over {(\lambda_0+2)^2\lambda_0^2}} \,{d\over dz} 
- \,{q^2\over {{4 z^2}}}\,{d^2\over dz^2}
\end{eqnarray*}

\begin{eqnarray*}
Q^{(3)}_{\ell=2}&=&
\left[\left({{i\over 2}\,m\,q + {{5\,{q^2}}\over 8} - 
        {{5\,{m^2}\,{q^2}}\over 9} + 
        {{11\,i}\over {24}}\,m\,{q^3} - 
        {{11\,i}\over {54}}\,{m^3}\,{q^3}}\right)
        {1\over {{z^4}}} \right.\\
&+& \left({{{-\left( m\,q \right) }\over 
          {24}} + {{5\,i}\over {48}}\,{q^2} - 
        {{65\,i}\over {216}}\,{m^2}\,{q^2} - 
        {{m\,{q^3}}\over 2} + 
        {{16\,{m^3}\,{q^3}}\over {81}}}\right)
        {1\over {{z^3}}} \\
&+&\left. \left({{i\over {24}}\,m\,q - 
        {{{q^2}}\over {48}} + 
        {{17\,{m^2}\,{q^2}}\over {216}} - 
        {{65\,i}\over {378}}\,m\,{q^3} + 
        {{17\,i}\over {252}}\,{m^3}\,{q^3}}\right){1\over 
      {{z^2}}} \right]
{d\over dz}\\
&+&
\left[ \left({i\,m\,q + {{7\,{q^2}}\over 4} - 
        {{25\,{m^2}\,{q^2}}\over {18}} + 
        {{4\,i}\over 3}\,m\,{q^3} - 
        {{29\,i}\over {54}}\,{m^3}\,{q^3}}\right)
        {1\over {{z^5}}} \right.\\
&+& \left({{{-5\,m\,q}\over {12}} + 
        {{19\,i}\over {24}}\,{q^2} - 
        {{103\,i}\over {108}}\,{m^2}\,{q^2} - 
        {{19\,m\,{q^3}}\over {12}} + 
        {{101\,{m^3}\,{q^3}}\over {162}}}\right)
        {1\over {{z^4}}} \\
&+& \left({{{-{q^2}}\over 8} + 
        {{41\,{m^2}\,{q^2}}\over {108}} - 
        {{127\,i}\over {189}}\,m\,{q^3} + 
        {{601\,i}\over {2268}}\,{m^3}\,{q^3}}\right){1\over 
      {{z^3}}} \\
&+&\left.\left( {{{-\left( m\,q \right) }\over 
          {24}} - {i\over {48}}\,{q^2} + 
        {{17\,i}\over {216}}\,{m^2}\,{q^2} + 
        {{\lambda_3\,{q^3}}\over 8} + 
        {{65\,m\,{q^3}}\over {378}} - 
        {{17\,{m^3}\,{q^3}}\over {252}}}\right){1\over 
      {{z^2}}} \right]\,,
\end{eqnarray*}

\begin{eqnarray*}
Q^{(3)}_{\ell=3}&=&
\left[\left({i\over {10}}\,m\,q + {{{q^2}}\over 8} - 
        {{7\,{m^2}\,{q^2}}\over {180}} + 
        {i\over {60}}\,m\,{q^3} - 
        {{7\,i}\over {540}}\,{m^3}\,{q^3}\right)
        {1\over {{z^4}}} \right.\\
&+&\left({{{-7\,m\,q}\over {600}} - 
        {{83\,i}\over {1200}}\,{q^2} - 
        {{7\,i}\over {1350}}\,{m^2}\,{q^2} - 
        {{23\,m\,{q^3}}\over {720}} + 
        {{19\,{m^3}\,{q^3}}\over {3240}}}\right)
        {1\over {{z^3}}} \\
&+&\left.\left({{i\over {1200}}\,m\,q - 
        {{7\,{q^2}}\over {1200}} + 
        {{23\,{m^2}\,{q^2}}\over {5400}} - 
        {{7\,i}\over {2160}}\,m\,{q^3} + 
        {{11\,i}\over {12960}}\,{m^3}\,{q^3}}\right)
        {1\over {{z^2}}} \right]{d\over dz}\\
&+&
\left[\left({{i\over 5}\,m\,q + {{{q^2}}\over 2} - 
        {{31\,{m^2}\,{q^2}}\over {180}} + 
        {{23\,i}\over {120}}\,m\,{q^3} - 
        {{5\,i}\over {108}}\,{m^3}\,{q^3}}\right){1\over 
      {{z^5}}} \right.\\
&+& \left({{{-8\,m\,q}\over {75}} + 
        {{109\,i}\over {1200}}\,{q^2} - 
        {{79\,i}\over {1350}}\,{m^2}\,{q^2} - 
        {{25\,m\,{q^3}}\over {144}} + 
        {{31\,{m^3}\,{q^3}}\over {810}}}\right){1\over 
      {{z^4}}} \\
&+&\left( {{{-13\,i}\over {1200}}\,m\,q + 
        {{19\,{q^2}}\over {300}} + 
        {{17\,{m^2}\,{q^2}}\over {1800}} - 
        {{19\,i}\over {540}}\,m\,{q^3} + 
        {{29\,i}\over {4320}}\,{m^3}\,{q^3}}\right)
        {1\over {{z^3}}} \\
&+&\left. \left({{{-\left( m\,q \right) }\over 
          {1200}} - {{7\,i}\over {1200}}\,{q^2} + 
        {{23\,i}\over {5400}}\,{m^2}\,{q^2} + 
        {{\lambda_3\,{q^3}}\over 8} + 
        {{7\,m\,{q^3}}\over {2160}} - 
        {{11\,{m^3}\,{q^3}}\over {12960}}}\right)
        {1\over {{z^2}}} \right].
\end{eqnarray*}

\begin{eqnarray*}
Q^{(4)}_{\ell=2}&=&
\left[\left({{{-3\,{q^2}}\over 8} + 
        {{3\,{m^2}\,{q^2}}\over 8} - 
        {{23\,i}\over {24}}\,m\,{q^3} + 
        {{43\,i}\over {108}}\,{m^3}\,{q^3} - 
        {{5\,{q^4}}\over {32}} + 
        {{4\,{m^2}\,{q^4}}\over 9} - 
        {{79\,{m^4}\,{q^4}}\over {648}}}\right){1\over 
      {{z^5}}} \right.\\
&+&\left({{-\left( m\,q \right) }\over 
          8} - {{3\,i}\over {32}}\,{q^2} + 
        {{5\,i}\over {72}}\,{m^2}\,{q^2} + 
        {{77\,m\,{q^3}}\over {96}} - 
        {{263\,{m^3}\,{q^3}}\over {648}} - \right.\\
& &\left.
        {{19\,i}\over {96}}\,{q^4} + 
        {{79\,i}\over {108}}\,{m^2}\,{q^4} - 
        {{137\,i}\over {648}}\,{m^4}\,{q^4}\right)
        {1\over {{z^4}}} \\
&+&\left({{-i}\over {96}}\,m\,q - 
        {{5\,{q^2}}\over {192}} + 
        {{73\,{m^2}\,{q^2}}\over {864}} + 
        {{13\,i}\over {108}}\,m\,{q^3} - 
        {i\over 9}\,{m^3}\,{q^3} + \right.\\
& &\left.
        {{25\,{q^4}}\over {252}} - 
        {{109\,{m^2}\,{q^4}}\over {252}} + 
        {{680\,{m^4}\,{q^4}}\over {5103}}\right){1\over 
      {{z^3}}} \\
&+&\left({{-\left( m\,q \right) }\over 
          {96}} - {i\over {192}}\,{q^2} + 
        {{11\,i}\over {288}}\,{m^2}\,{q^2} + 
        {{5\,m\,{q^3}}\over {108}} - 
        {{{m^3}\,{q^3}}\over {3888}} + 
        {{7\,i}\over {432}}\,{q^4} + \right.\\
& &\left.\left.
        {i\over {72}}\,\lambda_3\,m\,{q^4} - 
        {{1349\,i}\over {13608}}\,{m^2}\,{q^4} + 
        {{509\,i}\over {15309}}\,{m^4}\,{q^4}\right){1\over 
      {{z^2}}} \right]{d\over dz}\\
&+&
\left[\left({{{-3\,{q^2}}\over 4} + 
        {{7\,{m^2}\,{q^2}}\over {12}} - 
        {{47\,i}\over {24}}\,m\,{q^3} + 
        {{19\,i}\over {27}}\,{m^3}\,{q^3} - 
        {{7\,{q^4}}\over {16}} + 
        {{65\,{m^2}\,{q^4}}\over {72}} - 
        {{71\,{m^4}\,{q^4}}\over {324}}}\right){1\over 
      {{z^6}}} \right.\\
&+&\left({{-\left( m\,q \right) }\over 
          4} - {{7\,i}\over {16}}\,{q^2} + 
        {{3\,i}\over 8}\,{m^2}\,{q^2} + 
        {{29\,m\,{q^3}}\over {12}} - 
        {{659\,{m^3}\,{q^3}}\over {648}} - \right.\\
& &\left.
        {{29\,i}\over {48}}\,{q^4} + 
        {{49\,i}\over {27}}\,{m^2}\,{q^4} - 
        {{53\,i}\over {108}}\,{m^4}\,{q^4}\right){1\over 
      {{z^5}}} \\
&+&\left({{-5\,i}\over {48}}\,m\,q + 
        {{5\,{q^2}}\over {96}} + 
        {{41\,{m^2}\,{q^2}}\over {432}} + 
        {{2287\,i}\over {3024}}\,m\,{q^3} - 
        {{2039\,i}\over {4536}}\,{m^3}\,{q^3} + \right.\\
& &\left.
        {{101\,{q^4}}\over {288}} - 
        {{3991\,{m^2}\,{q^4}}\over {3024}} + 
        {{15853\,{m^4}\,{q^4}}\over {40824}}\right)
        {1\over {{z^4}}} \\
&+&\left({{-i}\over {32}}\,{q^2} + 
        {{53\,i}\over {432}}\,{m^2}\,{q^2} - 
        {{2\,m\,{q^3}}\over {27}} + 
        {{431\,{m^3}\,{q^3}}\over {3888}} + 
        {{349\,i}\over {3024}}\,{q^4} + \right.\\
& &\left.
        {i\over {72}}\,\lambda_3\,m\,{q^4} - 
        {{7235\,i}\over {13608}}\,{m^2}\,{q^4} + 
        {{2549\,i}\over {15309}}\,{m^4}\,{q^4}\right)
        {1\over  {{z^3}}} \\
&+&\left({{-i}\over {96}}\,m\,q + 
        {{{q^2}}\over {192}} - 
        {{11\,{m^2}\,{q^2}}\over {288}} + 
        {{5\,i}\over {108}}\,m\,{q^3} - 
        {i\over {3888}}\,{m^3}\,{q^3} - 
        {{7\,{q^4}}\over {432}} + 
        {{\lambda_4\,{q^4}}\over {16}} - \right.\\
& &\left.\left.
        {{\lambda_3\,m\,{q^4}}\over {72}} + 
        {{1349\,{m^2}\,{q^4}}\over {13608}} - 
        {{509\,{m^4}\,{q^4}}\over {15309}}\right){1\over 
      {{z^2}}} \right]. 
\end{eqnarray*}

\section{$X^{\lowercase{\rm in}}_{\ell \lowercase{m} \omega}$}
\noindent
(a)$\ell=2$

\begin{eqnarray*}
X^{{\rm in}}_{2 m\omega}&=&
{{{z^3}}\over {15}} - {{{z^5}}\over {210}} + {{{z^7}}\over {7560}} - 
  {{{z^9}}\over {498960}} + {{{z^{11}}}\over {51891840}}\\
&+&\epsilon\left(
{i\over {30}}\,m\,q\,{z^2} - {{13\,{z^4}}\over {630}} - 
  {{11\,i}\over {3780}}\,m\,q\,{z^4} + {{{z^6}}\over {810}} + 
  {{13\,i}\over {136080}}\,m\,q\,{z^6} \right.\\
& &\left.
- {{53\,{z^8}}\over {1782000}} -  {i\over {598752}}\,m\,q\,{z^8}\right)\\
&+&\epsilon^2\left(
\left( {i\over {60}}\,m\,q + {{{q^2}}\over {120}} - 
     {{{m^2}\,{q^2}}\over {120}} \right) \,z - {{m\,q\,{z^2}}\over {30}} 
\right.\\
& &
+ {z^3}\,\left( {{26743}\over {110250}} - {{433\,i}\over {22680}}\,m\,q - 
     {{3\,{q^2}}\over {3920}} + {{79\,{m^2}\,{q^2}}\over {105840}} - 
     {{107\,\ln z}\over {3150}} \right)  \\
& &
+ \left( {{m\,q}\over {270}} - {i\over {7560}}\,{q^2} + 
     {i\over {34020}}\,{m^2}\,{q^2} \right) \,{z^4} \\
& &\left.
+ {z^5}\,\left( -{{140953}\over {9261000}} + {{17\,i}\over {12960}}\,m\,q + 
     {{11\,{q^2}}\over {423360}} - {{19\,{m^2}\,{q^2}}\over {762048}} + 
     {{107\,\ln z}\over {44100}} \right) 
\right)\\
&+&\epsilon^3\left(
{{-i}\over {90}}\,m\,q - {{{q^2}}\over {30}} + {{{m^2}\,{q^2}}\over {40}} - 
  {{19\,i}\over {720}}\,m\,{q^3} + {{7\,i}\over {720}}\,{m^3}\,{q^3} 
\right.\\
& &
+ \left( {{-\left( m\,q \right) }\over {36}} - {i\over {120}}\,{q^2} + 
     {{m\,{q^3}}\over {36}} - {{{m^3}\,{q^3}}\over {90}} \right) \,z \\
& &
+ z^2 \left( {{319}\over {6300}} + {{2074\,i}\over {18375}}\,m\,q - 
     {{41\,{q^2}}\over {5040}} + {{{m^2}\,{q^2}}\over {648}} + 
     {{2887\,i}\over {211680}}\,m\,{q^3} 
\right.\\
& &\left.\left.
-    {{1153\,i}\over {211680}}\,{m^3}\,{q^3} - 
     {{107\,i}\over {6300}}\,m\,q\,\ln z \right) 
\right)\\
&+&\epsilon^4\left(
{{-i}\over {120}}\,m\,q + {{17\,{m^2}\,{q^2}}\over {1440}} + 
     {{11\,i}\over {480}}\,m\,{q^3} - {i\over {480}}\,{m^3}\,{q^3} + 
     {{23\,{q^4}}\over {1920}} \right.\\
& &\left.
- {{11\,{m^2}\,{q^4}}\over {576}} + 
     {{17\,{m^4}\,{q^4}}\over {5760}}
\right){1\over z}.
\end{eqnarray*}

\noindent
(b)$\ell=3$
\begin{eqnarray*}
X^{{\rm in}}_{3 m\omega}&=&
{{{z^4}}\over {105}} - {{{z^6}}\over {1890}} + {{{z^8}}\over {83160}} - 
  {{{z^{10}}}\over {6486480}}\\
&+&\epsilon\left(
{{-{z^3}}\over {126}} + {{2\,i}\over {945}}\,m\,q\,{z^3} - 
  {{{z^5}}\over {630}} - {i\over {7560}}\,m\,q\,{z^5} + 
  {{221\,{z^7}}\over {2494800}} + {i\over {299376}}\,m\,q\,{z^7}
\right)\\
&+&\epsilon^2\left(
\left( {{-i}\over {630}}\,m\,q + {{{q^2}}\over {840}} + 
     {{{m^2}\,{q^2}}\over {7560}} \right) \,{z^2} + 
  \left( {{-m\,q }\over {540}} - {i\over {1512}}\,{q^2} + 
     {i\over {6804}}\,{m^2}\,{q^2} \right) \,{z^3} 
\right.\\
& &\left.
+  {z^4}\,\left( {{76369}\over {1852200}} - {{299\,i}\over {453600}}\,m\,q - 
     {{{q^2}}\over {10800}} - {{{m^2}\,{q^2}}\over {75600}} - 
     {{13\,\ln z}\over {4410}} \right) 
\right)\\
&+&\epsilon^3\left(
\left( {{-i}\over {2520}}\,m\,q - {{{q^2}}\over {1008}} + 
    {{{m^2}\,{q^2}}\over {2160}} - {i\over {7560}}\,m\,{q^3} + 
    {i\over {7560}}\,{m^3}\,{q^3} \right) \,z
\right).
\end{eqnarray*}

\noindent
(c)$\ell=4$
\begin{eqnarray*}
X^{{\rm in}}_{4 m\omega}&=&
{{{z^5}}\over {945}} - {{{z^7}}\over {20790}} + {{{z^9}}\over {1081080}}\\
&+&\epsilon\left(
{{-{z^4}}\over {630}} + {i\over {7560}}\,m\,q\,{z^4} -
 {{{z^6}}\over {9900}} - 
  {i\over {154000}}\,m\,q\,{z^6}
\right)\\
&+&\epsilon^2\left(
{{{z^3}}\over {1764}} - {i\over {5040}}\,m\,q\,{z^3} + 
  {{{q^2}\,{z^3}}\over {4410}} + {{{m^2}\,{q^2}\,{z^3}}\over {52920}}
\right).
\end{eqnarray*}

\noindent
(d)$\ell=5$
\begin{eqnarray*}
X^{{\rm in}}_{5 m\omega}&=&
{{{z^6}}\over {10395}} - {{{z^8}}\over {270270}}
+\epsilon\left(
{{-{z^5}}\over {4950}} + {{2\,i}\over {259875}}\,m\,q\,{z^5}
\right).
\end{eqnarray*}

\noindent
(d)$\ell=6$
\begin{eqnarray*}
X^{{\rm in}}_{6 m\omega}&=&{{{z^7}}\over {135135}}.
\end{eqnarray*}

\section{$R^{\lowercase{\rm in}}_{\ell \lowercase{m}\omega}$}
\noindent
(a)$\ell=2$
\begin{eqnarray*}
\omega R^{{\rm in}}_{2 m\omega}&=&
{{{z^4}}\over {30}} + {i\over {45}}\,{z^5} - {{11\,{z^6}}\over {1260}} - 
  {i\over {420}}\,{z^7} + {{23\,{z^8}}\over {45360}} +
  {i\over {11340}}\,{z^9} \\
& &
- {{13\,{z^{10}}}\over {997920}} - {i\over {598752}}\,{z^{11}} + 
  {{59\,{z^{12}}}\over {311351040}} \\
&+&\epsilon\left(
{{-{z^3}}\over {15}} - {i\over {60}}\,m\,q\,{z^3} - {i\over {60}}\,{z^4} + 
  {{m\,q\,{z^4}}\over {45}} - {{41\,{z^5}}\over {3780}} + 
  {{277\,i}\over {22680}}\,m\,q\,{z^5} 
\right.\\
& &
- {{31\,i}\over {3780}}\,{z^6} - 
  {{7\,m\,q\,{z^6}}\over {1620}} 
+ {{17\,{z^7}}\over {5670}} - 
  {{61\,i}\over {54432}}\,m\,q\,{z^7} \\
& &
\left.
+ {{41\,i}\over {54432}}\,{z^8} + 
  {{47\,m\,q\,{z^8}}\over {204120}} - {{1579\,{z^9}}\over {10692000}} + 
  {{703\,i}\over {17962560}}\,m\,q\,{z^9}
\right)\\
&+&\epsilon^2\left(
{{{z^2}}\over {30}} + {i\over {40}}\,m\,q\,{z^2} + 
{{{q^2}\,{z^2}}\over {60}} - 
  {{{m^2}\,{q^2}\,{z^2}}\over {240}} - {i\over {60}}\,{z^3} - 
  {{m\,q\,{z^3}}\over {30}} + {i\over {90}}\,{q^2}\,{z^3} 
\right.\\
& & 
- {i\over {120}}\,{m^2}\,{q^2}\,{z^3} 
+ {{7937\,{z^4}}\over {55125}} - 
  {{53\,i}\over {9072}}\,m\,q\,{z^4} - {{101\,{q^2}\,{z^4}}\over {35280}} + 
  {{4213\,{m^2}\,{q^2}\,{z^4}}\over {635040}} \\
& &
+ {{4673\,i}\over {55125}}\,{z^5} - {{13\,m\,q\,{z^5}}\over {2835}} - 
  {{5\,i}\over {63504}}\,{q^2}\,{z^5} + 
  {{3503\,i}\over {1143072}}\,{m^2}\,{q^2}\,{z^5} 
- {{1665983\,{z^6}}\over {55566000}} \\
& &
- {{1777\,i}\over {544320}}\,m\,q\,{z^6}
- {{{q^2}\,{z^6}}\over {5040}} - 
  {{643\,{m^2}\,{q^2}\,{z^6}}\over {653184}}
- {{107\,{z^4}\,\ln z}\over {6300}} \\
& &\left.
- {{107\,i}\over {9450}}\,{z^5}\,\ln z + 
  {{1177\,{z^6}\,\ln z}\over {264600}}
\right)\\
&+&\epsilon^3\left(
\left( {{-i}\over {180}}\,m\,q - {{{q^2}}\over {60}} + 
     {{{m^2}\,{q^2}}\over {240}} - {i\over {144}}\,m\,{q^3} + 
     {i\over {1440}}\,{m^3}\,{q^3} \right) \,z 
\right.\\
& &
+ \left( {i\over {120}} + {{2\,m\,q}\over {135}} - {i\over {360}}\,{q^2} + 
     {{19\,i}\over {1440}}\,{m^2}\,{q^2} + {{11\,m\,{q^3}}\over {1080}} - 
     {{{m^3}\,{q^3}}\over {540}} \right) \,{z^2} \\
& &
+ {z^3}\,\left( -{{10933}\over {49000}} - 
     {{578569\,i}\over {7938000}}\,m\,q - 
     {{677\,{q^2}}\over {52920}} - 
     {{529\,{m^2}\,{q^2}}\over {63504}} 
\right.\\
& &\left.\left.
   + {{317\,i}\over {63504}}\,m\,{q^3} - 
     {{167\,i}\over {84672}}\,{m^3}\,{q^3} + 
     {{107\,\ln z}\over {3150}} + 
     {{107\,i}\over {12600}}\,m\,q\,\ln z
      \right) 
\right)\\
&+&\epsilon^4\left(
{{-i}\over {720}}\,m\,q + {{{m^2}\,{q^2}}\over {2880}} + 
  {i\over {288}}\,m\,{q^3} - {i\over {2880}}\,{m^3}\,{q^3} + 
  {{{q^4}}\over {480}} - {{{m^2}\,{q^4}}\over {720}} + 
  {{{m^4}\,{q^4}}\over {11520}}
\right).
\end{eqnarray*}

\noindent
(b)$\ell=3$
\begin{eqnarray*}
\omega R^{{\rm in}}_{3 m\omega}&=&
  {{{z^5}}\over {630}} + {i\over {1260}}\,{z^6} -
  {{{z^7}}\over {3780}} - 
  {i\over {16200}}\,{z^8} + {{29\,{z^9}}\over {2494800}} \\
& &
+ {i\over {554400}}\,{z^{10}} - {{47\,{z^{11}}}\over {194594400}} \\
&+&\epsilon\left(
  {{-{z^4}}\over {252}} - {i\over {1890}}\,m\,q\,{z^4} -
  {i\over {756}}\,{z^5} + 
  {{11\,m\,q\,{z^5}}\over {22680}} + 
  {{19\,i}\over {90720}}\,m\,q\,{z^6} 
\right.\\
& &\left.
- {i\over {9450}}\,{z^7} - 
  {{m\,q\,{z^7}}\over {16200}} + 
  {{647\,{z^8}}\over {14968800}} - 
  {{247\,i}\over {17962560}}\,m\,q\,{z^8} 
\right)\\
&+&\epsilon^2\left(
{{{z^3}}\over {315}} + {i\over {945}}\,m\,q\,{z^3} + 
  {{{q^2}\,{z^3}}\over {1260}} - {{{m^2}\,{q^2}\,{z^3}}\over {15120}} + 
  {i\over {2520}}\,{z^4} - {{17\,m\,q\,{z^4}}\over {15120}} 
\right.\\
& &
+ {i\over {2160}}\,{q^2}\,{z^4} - 
  {{31\,i}\over {272160}}\,{m^2}\,{q^2}\,{z^4} + 
  {{81409\,{z^5}}\over {11113200}} \\
& &\left.
- {{313\,i}\over {907200}}\,m\,q\,{z^5} - 
  {{41\,{q^2}\,{z^5}}\over {226800}} + 
  {{617\,{m^2}\,{q^2}\,{z^5}}\over {8164800}} - 
  {{13\,{z^5}\,\ln z}\over {26460}}
\right)\\
&+&\epsilon^3\left(
  {{-{z^2}}\over {1260}} - {i\over {1680}}\,m\,q\,{z^2} - 
  {{{q^2}\,{z^2}}\over {840}} + {{{m^2}\,{q^2}\,{z^2}}\over {10080}} - 
  {i\over {5040}}\,m\,{q^3}\,{z^2}
\right).
\end{eqnarray*}

\noindent
(c)$\ell=4$
\begin{eqnarray*}
\omega R^{{\rm in}}_{4 m\omega}&=&
{{{z^6}}\over {11340}} + {i\over {28350}}\,{z^7} -
  {{13\,{z^8}}\over {1247400}} - {i\over {467775}}\,{z^9} +
  {{71\,{z^{10}}}\over {194594400}}\\
&+&\epsilon\left(
{{-{z^5}}\over {3780}} - {i\over {45360}}\,m\,q\,{z^5} -
  {{11\,i}\over {136080}}\,{z^6} + {{m\,q\,{z^6}}\over {64800}} 
\right. \\
& &\left.
+ {{131\,{z^7}}\over {18711000}} + {{697\,i}\over
{124740000}}\,m\,q\,{z^7}
\right)\\
&+&\epsilon^2\left(
{{{z^4}}\over {3528}} + {i\over {18144}}\,m\,q\,{z^4} +
{{{q^2}\,{z^4}}\over {21168}} 
- {{{m^2}\,{q^2}\,{z^4}}\over{635040}}
\right).\\
\end{eqnarray*}

\noindent
(d)$\ell=5$
\begin{eqnarray*}
\omega R^{{\rm in}}_{5 m\omega}&=&
{{{z^7}}\over {207900}} + {i\over {623700}}\,{z^8} - 
  {{{z^9}}\over {2316600}}\\
&+&\epsilon\left(
{{-{z^6}}\over {59400}} - {i\over {1039500}}\,m\,q\,{z^6}
\right).
\end{eqnarray*}

\noindent
(e)$\ell=6$
\begin{eqnarray*}
\omega R^{{\rm in}}_{6 m\omega}&=&
{{{z^8}}\over {4054050}}.
\end{eqnarray*}

\section{Spheroidal harmonics}

In this appendix, we describe the expansion of the spheroidal 
harmonics $_{-2}S^{a\omega}_{\ell m}$ at order $O((a\omega)^2)$. 

The spheroidal harmonics of spin weight $s=-2$ obey the equation,
\begin{eqnarray}
& &\Bigl[{1 \over \sin\theta}{d \over d\theta}
 \Bigl\{\sin\theta {d \over d\theta} \Bigr\}
-a^2\omega^2\sin^2\theta
 -{(m-2\cos\theta)^2 \over \sin^2\theta}
\nonumber\\
& &~~~~~~~~~~~~~~~
+4a\omega\cos\theta-2+2ma\omega+\lambda\Bigr] 
{}_{-2}S_{\ell m}^{a\omega}=0. \label{eq:spheroid}
\end{eqnarray}
We expand $_{-2}S_{\ell m}^{a\omega}$ and $\lambda$ as
\begin{eqnarray} 
{}_{-2}S_{\ell m}^{a\omega}
&=&{}_{-2}P_{\ell m}+a\omega S_{\ell m}^{(1)}
+(a\omega)^2 S_{\ell m}^{(2)}
+O((a\omega)^3), \nonumber\\
\lambda&=&\lambda_0(\ell)+a\omega\lambda_1(\ell)
+a^2\omega^2 \lambda_2(\ell)+O((a\omega)^3), \label{eq:slmexp2}
\end{eqnarray} 
where ${}_{-2}P_{\ell m}$ are the spherical harmonics of spin weight
$s=-2$ and $\lambda_n$ are given in section 2.2. 
Here we explicitly represent the $\ell$-dependence of $\lambda_n$
for later convenience. 
We set the normalization of ${}_{-2}P_{\ell m}$ as
\begin{equation}
\int_0^{\pi} |_{-2}P_{\ell m}|^2 \sin\theta d\theta=1.
\end{equation}

Inserting Eq.(\ref{eq:slmexp2}) into Eq.(\ref{eq:spheroid}) and
collecting the terms of order $(a\omega)^2$, we obtain
\begin{equation}
{\cal L}_0 S_{\ell m}^{(2)}+\lambda_0(\ell) S^{(2)}_{\ell m}
=-(4\cos\theta+2m+\lambda_1(\ell))S_{\ell m}^{(1)}
-(\lambda_2(\ell)-\sin^2\theta)~_{-2}P_{\ell m}\,,\label{eq:slm2}
\end{equation}
where ${\cal L}_0$ is the operator for the spin-weighted spherical 
harmonics,
\begin{eqnarray}
{\cal L}_0[{}_{-2}P_{\ell m}]
&\equiv&\left[ {1 \over \sin\theta}{d \over d\theta}
\Bigl\{ \sin\theta {d \over d\theta} \Bigr\}
-{(m-2\cos\theta)^2 \over \sin^2\theta}-2\right]
{}_{-2}P_{\ell m} \nonumber\\
&=&-\lambda_0~{}_{-2}P_{\ell m}\,.\\
\end{eqnarray}
By setting
\begin{eqnarray}
S_{\ell m}^{(1)}&=&\sum_{\ell'} c_{\ell m}^{\ell'}~{}_{-2}P_{\ell'm}\,,
\nonumber\\
S_{\ell m}^{(2)}&=&\sum_{\ell'} d_{\ell m}^{\ell'}~{}_{-2}P_{\ell'm}\,,
\end{eqnarray}
we insert it into Eq.(\ref{eq:slm2}), multiply it by 
${}_{-2}P_{\ell' m}$ and integrate it over $\theta$. 
Then we have 
\begin{eqnarray}
d_{\ell m}^{\ell'}&=&{1\over {\lambda_0(\ell)-\lambda_0(\ell')}}
\left[-\left(2m+\lambda_1(\ell)\right)
\left(c^{\ell+1}_{\ell m}\delta_{\ell', \ell+1}
+c^{\ell-1}_{\ell m}\delta_{\ell', \ell-1}\right)
-\delta_{\ell', \ell}\lambda_2(\ell)  
\right.\nonumber\\
& &
-4 c^{\ell+1}_{\ell m}\int d(\cos\theta) 
{}_{-2}P_{\ell' m}{}_{-2}P_{\ell+1 m} \cos\theta
-4 c^{\ell-1}_{\ell m}\int d(\cos\theta) 
{}_{-2}P_{\ell' m}{}_{-2}P_{\ell-1 m} \cos\theta
\nonumber\\
& &
\left.
+\int d(\cos\theta) 
{}_{-2}P_{\ell' m}{}_{-2}P_{\ell m} \sin^2\theta 
\right].
\end{eqnarray}
The integrals in this equation are given by \cite{Ref:press,Ref:cam} 
\begin{eqnarray*}
\int d(\cos\theta){}_{-2}P_{\ell' m}{}_{-2}P_{\ell m}\,\cos\theta
&=&\sqrt{{2\ell+1}\over {2\ell'+1}}
<\ell,1,m,0|\ell',m><\ell,1,2,0|\ell',2>, \\
\int d(\cos\theta){}_{-2}P_{\ell' m}{}_{-2}P_{\ell m}\,\sin^2\theta
&=&{2\over 3}\delta_{\ell',\ell} \\
& &-{2\over 3}\sqrt{{2\ell+1}\over {2\ell'+1}}
<\ell,2,m,0|\ell',m><\ell,2,2,0|\ell',2>, \\
\end{eqnarray*} 
where $<j_1,j_2,m_1,m_2|J,M>$ is a Clebsch-Gordan coefficient. 
Then, for $\ell=2$ and $3$, we obtain $d^{\ell'}_{\ell m}$ 
$(\ell'\neq\ell)$ which are given in section II.B. 
As for $d^{\ell}_{\ell m}$, we consider the normalization of
${}_{-2}P_{\ell m}$ (Eq.(\ref{eq:Snorm})). Inserting 
Eq.(\ref{eq:slmexp2}) into Eq.(\ref{eq:Snorm}), and
using the orthogonality of ${}_{-2}P_{\ell m}$, we obtain
\begin{eqnarray*} 
1&=&\int^\pi_0d\theta \sin\theta |{}_{-2}S_{\ell m}|^2\\ 
 &=&\int^\pi_0d\theta \sin\theta 
 \left\{({}_{-2}P_{\ell m})^2+2a\omega\sum_{\ell'}
c^{\ell'}_{\ell m}{}_{-2}P_{\ell' m}{}_{-2}P_{\ell m}
\right.\\
&+&
(a\omega)^2\sum_{\ell'\ell''}c^{\ell'}_{\ell m}c^{\ell''}_{\ell m}
{}_{-2}P_{\ell' m}{}_{-2}P_{\ell'' m} \\
&+&\left.  2(a\omega)^2\sum_{\ell'}
d^{\ell'}_{\ell m}{}_{-2}P_{\ell' m}{}_{-2}P_{\ell m}
+O\left((a\omega)^3\right)\right\} \\
&=&
1+(a\omega)^2\sum_{\ell'} \left(c^{\ell'}_{\ell m}\right)^2
+2(a\omega)^2d^\ell_{\ell m}+O((a\omega)^3).
\end{eqnarray*}
Then we have 
\begin{equation}
d^\ell_\ell=-{1\over 2}\left\{
\left(c^{\ell+1}_{\ell m}\right)^2+\left(c^{\ell-1}_{\ell m}\right)^2
\right\}.
\end{equation}

\section{The expression of the luminosity by means of the orbital angular 
frequency}

For the sake of convenience to calculate the 
orbital phase error, we describe the formula of gravitational wave luminosity 
by means of $v'\equiv (M\Omega)^{1/3}$. 
In this appendix, we define $\eta_{\ell, m}$ as
\begin{equation}
\left( {dE \over dt} \right)_{\ell m}
\equiv{16\over 5}\left(\mu\over M\right)^2
v'^{10}\eta_{\ell,m}. 
\nonumber 
\end{equation}
\begin{eqnarray}
\eta_{2,2}&=&
1 - {{107\,{v'^2}}\over {21}} + 
      \left( 4\,\pi  - {{8\,q}\over 3} \right) \,{v'^3} + 
      \left( {{4784}\over {1323}} + 2\,{q^2} \right) \,{v'^4} 
\nonumber\\
&+&   \left( {{-428\,\pi }\over {21}} + {{52\,q}\over {27}} \right) \,
       {v'^5} 
\nonumber\\
&+& \left( {{99210071}\over {1091475}} - 
         {{1712\,\gamma }\over {105}} + {{16\,{{\pi }^2}}\over 3} - 
         {{32\,\pi \,q}\over 3} - {{1817\,{q^2}}\over {567}} 
\right.
\nonumber\\
&-&\left.
  {{3424\,\ln 2}\over {105}} - {{1712\,\ln v'}\over {105}}
          \right)\,{v'^6}
+ \left( {{19136\,\pi }\over {1323}} 
+  {{364856\,q}\over {11907}} + 8\,\pi \,{q^2} - {{8\,{q^3}}\over 3}
          \right) \,{v'^7} 
\nonumber\\
&+& \left( -{{27956920577}\over {81265275}} + 
         {{183184\,\gamma }\over {2205}} - {{1712\,{{\pi }^2}}\over {63}} + 
         {{208\,\pi \,q}\over {27}} 
\right.
\nonumber\\
&+&\left.
         {{105022\,{q^2}}\over {9261}} + {q^4} + 
         {{366368\,\ln 2}\over {2205}} + 
         {{183184\,\ln v'}\over {2205}} \right)\,{v'^8}\,.
\\
\eta_{2,1}&=&
{{{v'^2}}\over {36}} - {{q\,{v'^3}}\over {12}} + 
      \left( -{{17}\over {504}} + {{{q^2}}\over {16}} \right) \,
       {v'^4} 
\nonumber\\
&+& \left( {{\pi }\over {18}} + 
         {{215\,q}\over {9072}} \right) \,{v'^5} 
\nonumber\\
&+& \left( -{{2215}\over {254016}} - {{\pi \,q}\over 6} + 
         {{313\,{q^2}}\over {1512}} \right) \,{v'^6} 
\nonumber\\
&+& \left( {{-17\,\pi }\over {252}} - {{18127\,q}\over {190512}} + 
         {{\pi \,{q^2}}\over 8} - {{7\,{q^3}}\over {24}} \right) \,
       {v'^7} 
\nonumber\\
&+& \left( {{15707221}\over {26195400}} - {{107\,\gamma }\over {945}} + 
         {{{{\pi }^2}}\over {27}} + {{215\,\pi \,q}\over {4536}} 
\right.
\nonumber\\
&+& \left.
         {{44299\,{q^2}}\over {95256}} + {{{q^4}}\over {16}} - 
         {{107\,\ln 2}\over {945}} - {{107\,\ln v'}\over {945}}
          \right)\, v'^8\,.
\\
\eta_{3,3}&=&
     {{1215\,{v'^2}}\over {896}} - 
      {{1215\,{v'^4}}\over {112}} + 
      \left( {{3645\,\pi }\over {448}} - {{1215\,q}\over {224}} \right) \,
       {v'^5} 
\nonumber\\
&+& \left( {{243729}\over {9856}} + 
         {{3645\,{q^2}}\over {896}} \right) \,{v'^6} + 
      \left( {{-3645\,\pi }\over {56}} + {{41229\,q}\over {1792}} \right) \,
       {v'^7} 
\nonumber\\
&+& \left( {{25037019729}\over {125565440}} - 
         {{47385\,\gamma }\over {1568}} + {{3645\,{{\pi }^2}}\over {224}} - 
         {{3645\,\pi \,q}\over {112}} 
\right.
\nonumber\\
&-&\left.
         {{236925\,{q^2}}\over {14336}} - 
         {{47385\,\ln 2}\over {1568}} - {{47385\,\ln 3}\over {1568}} - 
         {{47385\,\ln v'}\over {1568}} \right)\,{v'^8}\,.
\\
\eta_{3,2}&=&{{5\,{v'^4}}\over {63}} - 
      {{40\,q\,{v'^5}}\over {189}} + 
      \left( -{{193}\over {567}} + {{80\,{q^2}}\over {567}} \right) \,
       {v'^6} 
\nonumber\\
&+& \left( {{20\,\pi }\over {63}} + 
         {{982\,q}\over {1701}} \right) \,{v'^7} + 
      \left( {{86111}\over {280665}} - {{160\,\pi \,q}\over {189}} + 
         {{80\,{q^2}}\over {189}} \right) \,{v'^8}\,.
\\
\eta_{3,1}&=&{{{v'^2}}\over {8064}} - {{{v'^4}}\over {1512}} + 
      \left( {{\pi }\over {4032}} - {{25\,q}\over {18144}} \right) \,
       {v'^5} 
\nonumber\\
&+& \left( {{437}\over {266112}} + 
         {{17\,{q^2}}\over {24192}} \right) \,{v'^6} + 
      \left( {{-\pi }\over {756}} + {{2257\,q}\over {435456}} \right) \,
       {v'^7} 
\nonumber\\
&+& \left( -{{1137077}\over {50854003200}} - 
         {{13\,\gamma }\over {42336}} + {{{{\pi }^2}}\over {6048}} 
\right.
\nonumber\\
&-&\left.
         {{25\,\pi \,q}\over {9072}} + {{12863\,{q^2}}\over {3483648}} - 
         {{13\,\ln 2}\over {42336}} - {{13\,\ln v'}\over {42336}}
          \right)\,{v'^8}\,.
\\
\eta_{4,4}&=& {{1280\,{v'^4}}\over {567}} - 
      {{151808\,{v'^6}}\over {6237}} + 
      \left( {{10240\,\pi }\over {567}} - {{20480\,q}\over {1701}} \right) \,
       {v'^7} 
\nonumber\\
&+& \left( {{560069632}\over {6243237}} + 
         {{5120\,{q^2}}\over {567}} \right) \,{v'^8}\,.
\\
\eta_{4,3}&=&{{729\,{v'^6}}\over {4480}} - 
      {{729\,q\,{v'^7}}\over {1792}} + 
      \left( -{{28431}\over {24640}} + {{3645\,{q^2}}\over {14336}} \right) \,
       {v'^8}\,.
\\
\eta_{4,2}&=&{{5\,{v'^4}}\over {3969}} - 
      {{437\,{v'^6}}\over {43659}} + 
      \left( {{20\,\pi }\over {3969}} - {{170\,q}\over {11907}} \right) \,
       {v'^7} 
\nonumber\\
&+& \left( {{7199152}\over {218513295}} + 
         {{200\,{q^2}}\over {27783}} \right) \,{v'^8}\,.
\\
\eta_{4,1}&=& {{{v'^6}}\over {282240}} - 
      {{q\,{v'^7}}\over {112896}} + 
      \left( -{{101}\over {4656960}} + {{5\,{q^2}}\over {903168}} \right) \,
       {v'^8}\,.
\\
\eta_{5,5}&=& {{9765625\,{{{\it v'}}^6}}\over {2433024}} - 
      {{2568359375\,{{{\it v'}}^8}}\over {47443968}}\,.
\\
\eta_{5,4}&=&{{4096\,{v'^8}}\over {13365}}\,.
\\
\eta_{5,3}&=& {{2187\,{{{\it v'}}^6}}\over {450560}} - 
      {{150903\,{{{\it v'}}^8}}\over {2928640}}\,.
\\
\eta_{5,2}&=&{{4\,{v'^8}}\over {40095}}\,.
\\
\eta_{5,1}&=&{{{v'^6}}\over {127733760}} - 
      {{179\,{v'^8}}\over {2490808320}}\,.
\\
\eta_{6,6}&=&
{{26244\,{v'^8}}\over {3575}}\,.
\\
\eta_{6,4}&=&{{131072\,{v'^8}}\over {9555975}}\,.
\\
\eta_{6,2}&=&{{4\,{v'^8}}\over {5733585}}\,.
\end{eqnarray}

In total, 
\begin{eqnarray}
\biggl< {dE \over dt} \biggr>&=&{32\over 5}\left(\mu\over M\right)v'^{10}
\left( 1 - {{1247\,{v'^2}}\over {336}} + 
      \left( 4\,\pi  - {{11\,q}\over 4} \right) \,{v'^3} 
\right.
\nonumber\\
&+&  \left( -{{44711}\over {9072}} + {{33\,{q^2}}\over {16}} \right) \,
       {v'^4} + \left( {{-8191\,\pi }\over {672}} - 
         {{59\,q}\over {16}} \right) \,{v'^5} 
\nonumber\\
&+& \left( {{6643739519}\over {69854400}} - {{1712\,\gamma }\over {105}} + 
         {{16\,{{\pi }^2}}\over 3} - {{65\,\pi \,q}\over 6} + 
         {{611\,{q^2}}\over {504}} - {{3424\,\ln 2}\over {105}}
\right.
\nonumber\\
&-&\left. {{1712\,\ln v'}\over {105}} \right)\,{v'^6} 
+ \left( {{-16285\,\pi }\over {504}} + {{162035\,q}\over {3888}} + 
         {{65\,\pi \,{q^2}}\over 8} - {{71\,{q^3}}\over {24}} \right) \,
       {v'^7} 
\nonumber\\
&+& \left( -{{323105549467}\over {3178375200}} + 
         {{232597\,\gamma }\over {4410}} - {{1369\,{{\pi }^2}}\over {126}} - 
         {{359\,\pi \,q}\over {14}} + {{22667\,{q^2}}\over {4536}} 
\right. 
\nonumber\\ 
&+&\left.\left. 
         {{17\,{q^4}}\over {16}} + {{39931\,\ln 2}\over {294}} - 
         {{47385\,\ln 3}\over {1568}} + 
         {{232597\,\ln  v'}\over {4410}} \right)\,{v'^8}\right). 
\end{eqnarray}

\vskip 5mm

\begin{center}
\large{\bf Figure Captions}
\end{center}

\vskip 3mm

\noindent
Figs.1(a-e) : Error of the post-Newtonian formulas as
a function of the Boyer-Lindquist coordinate radius $r$
for $6 \leq r/M \leq 100$ 
in the case $q=-0.9$, $-0.5$, $0$, 0.5 and $0.9$.
In each figure, open square, filled triangle, open triangle,
filled circle, and open circle denote the error of
2-PN, 2.5-PN, 3-PN, 3.5-PN and 4-PN formulas, respectively.

\noindent
Fig.2 : Error of the post-Newtonian formula of $q=0.9$ 
for $2.5 \leq r/M \leq 12$. 
Open square, filled triangle, open triangle,
filled circle, and open circle denote the error of
2-PN, 2.5-PN, 3-PN, 3.5-PN and 4-PN formulas, respectively.

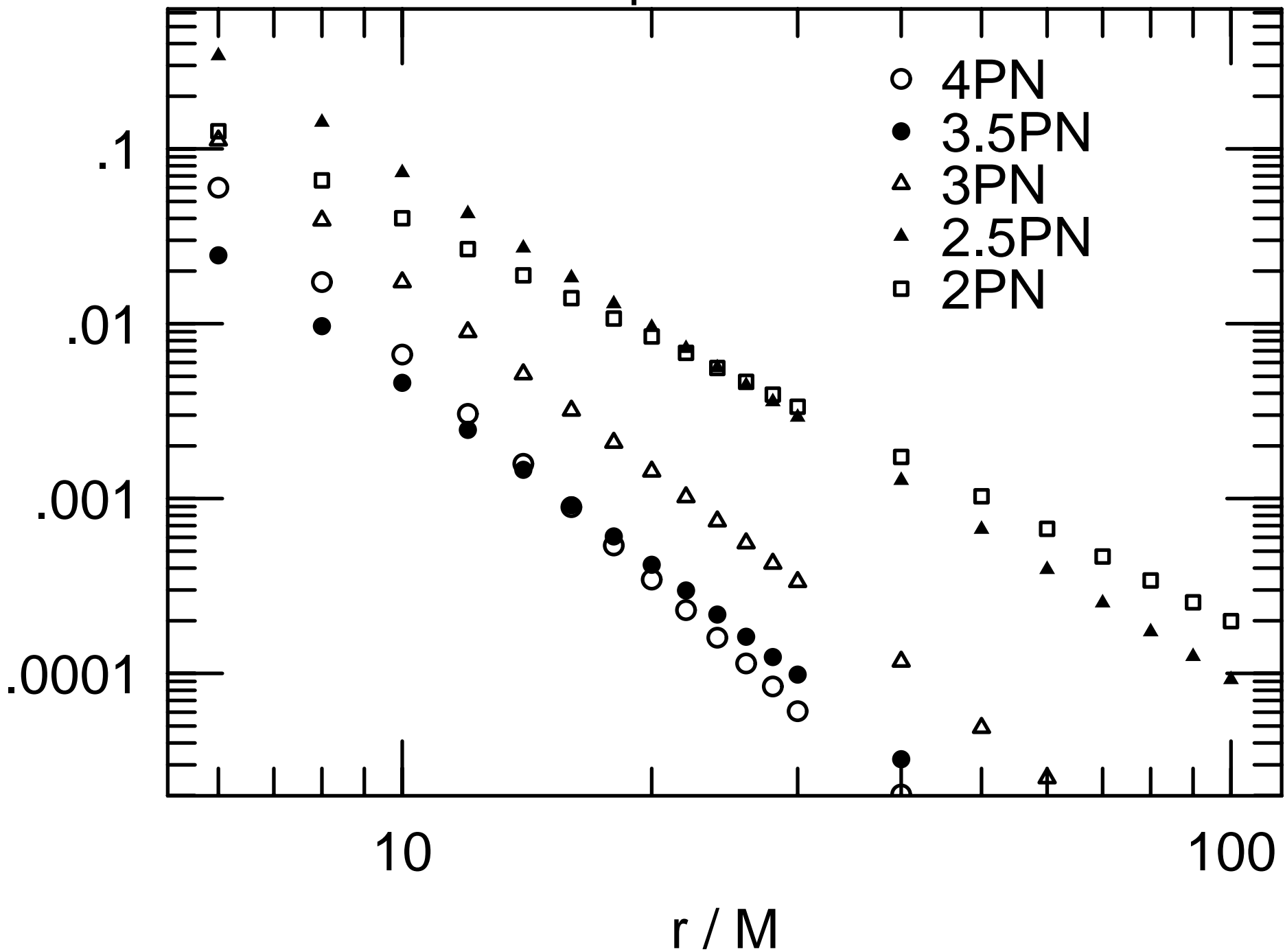
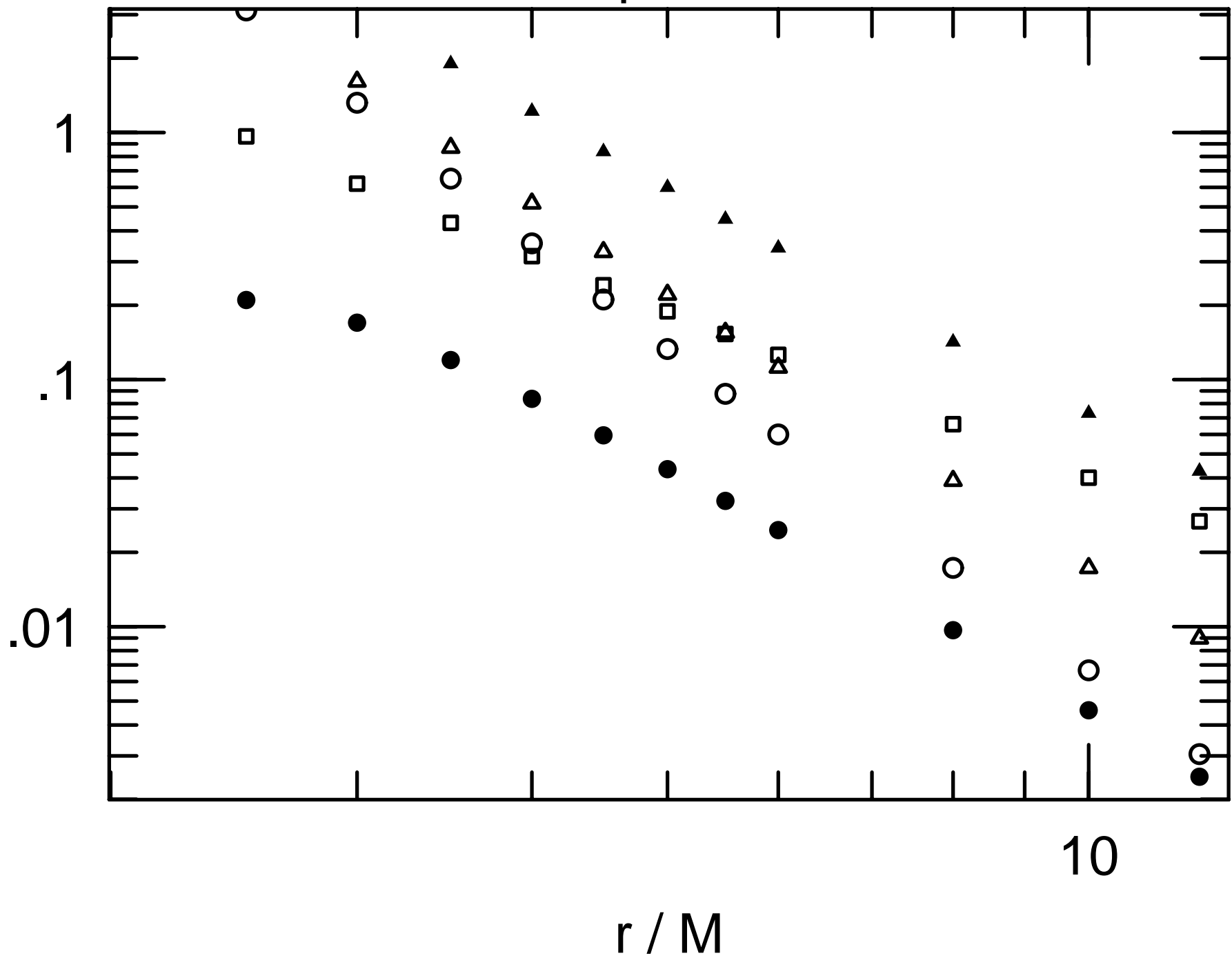
\end{document}